\documentclass[journal]{IEEEtran}
\usepackage{cite}
\ifCLASSINFOpdf
   \usepackage[pdftex]{graphicx}
  \graphicspath{{Figures/}}
  \DeclareGraphicsExtensions{.pdf,.jpeg,.png}
\else
\fi
 \usepackage[caption=false,font=footnotesize]{subfig}

\usepackage{soul,color}
\usepackage{silence}
\usepackage{amsmath}
\usepackage{amssymb}
\usepackage{float}
\usepackage{booktabs}
\usepackage{textcomp}
\usepackage{array}
\usepackage{mathtools}
\usepackage{glossaries}
\usepackage{multirow}
\usepackage{multicol}
\usepackage{amsmath}
\usepackage{bbm}
\usepackage{bm}
\usepackage[export]{adjustbox}
\usepackage[table]{xcolor}
\usepackage{bigstrut}
\usepackage{graphicx}
\graphicspath{{./Figures/}}
\newcommand{\Section}[1] {Section~#1}
\newcommand{\Figure}[1] {Fig.~#1}
\newcommand{\Table}[1] {Table~#1}

\newacronym{SDSCE}{SDSCE}{Simultaneous Double Stimulus for Continuous Evaluation}
\newacronym{DSCQS}{DSCQS}{Double Stimulus Continuous Quality Scale}
\newacronym[plural=BTCs, firstplural=Basic Test Cells (BTCs)]{BTC}{BTC}{Basic Test Cell}
\newacronym{HFR}{HFR}{High Frame-Rate}
\newacronym{HDR}{HDR}{High Dynamic Range}
\newacronym{UHD}{UHD}{Ultra-High Definition}
\newacronym{HD}{HD}{High Definition}
\newacronym[plural=RFs, firstplural=Random Forests (RFs)]{RF}{RF}{Random Forest}
\newacronym{VFR}{VFR}{Variable Frame-Rate}
\newacronym{ML}{ML}{Machine Learning}
\newacronym{FD}{FD}{Frame Decimation}
\newacronym{RFE}{RFE}{Recursive Feature Elimination}
\newacronym{HEVC}{HEVC}{High Efficiency Video Coding}
\newacronym{TP}{TP}{True Positives}
\newacronym{FP}{FP}{False Positives}
\newacronym{TN}{TN}{True Negatives}
\newacronym{FN}{FN}{False Negatives}
\newacronym{IBC}{IBC}{International Broadcasting Convention}
\newacronym{NAB}{NAB}{National Association of Broadcasters}
\newacronym{MVs}{MVs}{Motion Vectors}
\newacronym{CTC}{CTC}{Common Test Conditions}
\newacronym{GOP}{GOP}{Group Of Pictures}
\newacronym{POC}{POC}{Picture Order Count}
\newacronym[plural=TLs]{TL}{TL}{Temporal Layer}
\newacronym{QP}{QP}{Quantization Parameter}
\newacronym{RA}{RA}{Random Access}
\newacronym{UHDTV}{UHDTV}{Ultra-High Definition TV}
\newacronym{QoE}{QoE}{Quality of Experience}
\newacronym{HDTV}{HDTV}{High Definition TV}
\newacronym{DVB}{DVB}{Digital Video Broadcasting}
\newacronym{fps}{fps}{frames per second}
\newacronym{BBC}{BBC}{British Broadcasting Corporation}
\newacronym{ITU}{ITU}{International Telecommunication Union}
\newacronym[plural=CRTs,firstplural=Cathode Ray Tubes (CRTs)]{CRT}{CRT}{Cathode Ray Tube}
\newacronym{HVS}{HVS}{Human Visual System}
\newacronym{CSF}{CSF}{Contrast Sensitivity Function}
\newacronym{MDI}{MDI}{Mean Decrease Impurity}
\newacronym{SVR}{SVR}{Support Vector Regression}
\newacronym{GLCM}{GLCM}{Gray Level Co-occurrence Matrix}
\newacronym{OF}{OF}{Optical Flow}
\newacronym{MOS}{MOS}{Mean Opinion Score}
\newacronym{DMOS}{DMOS}{Differential Mean Opinion Score}
\newacronym{DCR}{DCR}{Degradation Category Rating}
\newacronym[plural=CIs]{CI}{CI}{Confidence Interval}
\newacronym{CNN}{CNN}{Convolutional Neural Network}
\newacronym{BVI-HFR}{BVI-HFR}{Bristol Vision Institute High Frame-Rate}

\begin{document}
\title{Quality-driven Variable Frame-Rate for Green Video Coding in Broadcast Applications}

\author{\hspace{4em} Glenn Herrou, Charles Bonnineau, Wassim Hamidouche,~\IEEEmembership{Member,~IEEE,} \newline Patrick Dumenil,  J\'er\^ome Fournier and Luce Morin% <-this % stops a space
\thanks{Manuscript received  April 13th, 2020; revised September 6th, 2020 and November 17th, 2020; accepted December 9th, 2020. This work has been funded by the French government through the ANR Investment referenced 10-AIRT-0007.}%
\thanks{All authors are with the Institute of Research and Technology (IRT) b\textless\textgreater com, 35510 Cesson S\'evign\'e, France.}% <-this % stops a space
\thanks{G. Herrou, C. Bonnineau, W. Hamidouche and L. Morin are also with INSA Rennes, Institut d'Electronique et des Technologies du Num\'eRique (IETR), CNRS - UMR 6164, VAADER team, 20 Avenue des Buttes de Coesmes, 35708 Rennes, France, (e-mail: glenn.herrou@insa-rennes.fr).}% <-this % stops a space
\thanks{C. Bonnineau is also with the Direction Technique, TDF, 35510 Cesson-\'evign\'e, France}%
\thanks{P. Dumenil is also with Harmonic Inc., 35510 Cesson-S\'evign\'e, France}%
\thanks{J. Fournier is also with Orange Labs, 35510 Cesson-S\'evign\'e, France}}%

%                 ^------------^------------^----Do not want these spaces!

\markboth{IEEE TRANSACTIONS ON CIRCUITS AND SYSTEMS FOR VIDEO TECHNOLOGY - ACCEPTED VERSION}%
{Herrou \MakeLowercase{\textit{et al.}}: Quality-driven Variable Frame-Rate for Green Video Coding in Broadcast Applications}

% \IEEEpubid{}


% \IEEEspecialpapernotice{(Invited Paper)}

\maketitle

\begin{abstract}
The Digital Video Broadcasting (DVB)  has proposed to introduce the Ultra-High Definition services in three phases: UHD-1 phase 1, UHD-1 phase 2 and UHD-2.  The UHD-1 phase 2 specification includes several new features such as High Dynamic Range (HDR) and High Frame-Rate (HFR).  It has been shown in several studies that HFR (+100 fps) enhances the perceptual quality and that this quality enhancement is content-dependent. On the other hand, HFR brings several challenges to the transmission chain including codec complexity increase and bit-rate overhead, which may delay or even prevent its deployment in the broadcast echo-system. In this paper, we propose a Variable Frame Rate (VFR) solution to determine the minimum (critical) frame-rate that preserves the perceived video quality of HFR video. The frame-rate determination is modeled as a 3-class classification problem which consists in dynamically and locally selecting one frame-rate among three: 30, 60 and 120 frames per second. Two random forests classifiers are trained with a ground truth carefully built by experts for this purpose. The subjective results conducted on ten HFR video contents, not included in the training set, clearly show the efficiency of the proposed solution enabling to locally determine the lowest possible frame-rate while preserving the quality of the HFR content. Moreover, our VFR solution enables significant bit-rate savings and complexity reductions at both encoder and decoder sides.
\end{abstract}


% \fi
% \IEEEpeerreviewmaketitle

%\hl{angular velocities of the main object can be a important criteria}
% \vfill
% \newpage

\section{Introduction}
\label{sec:intro}
\IEEEPARstart{T}{he deployment} of the latest \gls{UHDTV} system~\cite{itu-r_recommendation_BT2020-1} aims to increase the user's \gls{QoE} by introducing to the existing \gls{HDTV} system~\cite{itu-r_recommendation_BT709} new features such as higher spatial resolution, \gls{HDR}, wider color gamut and \gls{HFR}~\cite{nilsson2015ultra,UHDTV_IEEE}. Technical definition of the \gls{UHDTV} signal is available in the BT. 2020 recommendation of the \gls{ITU}~\cite{itu-r_recommendation_BT2020-1}. The \gls{UHD}-1 Phase 2 specification enables to increase the video frame-rate from 50/60 \gls{fps} in phase 1 to 100/120 \gls{fps}. 

Several papers~\cite{mackin2017high, kuroki2007psychophysical,noland2014application, laird2006spatio, hulusic2017quality, mackin2017investigating, salmon2013higher} have investigated the \gls{HFR}\footnote{In this paper high frame-rate video refers to video represented by 100 frames per second and more (+100 \gls{fps}).} video signal and shown its impact to enhance the viewing experience by reducing temporal artifacts specifically motion blur and temporal aliasing~\cite{mackin2017high}. Authors in~\cite{hulusic2017quality} have conducted subjective evaluations and shown that the combination of \gls{HFR} (100 \gls{fps}) and high resolution (4K) significantly increases the \gls{QoE} when the video is coded at high bit-rate. The subjective evaluations conducted in~\cite{mackin2017investigating} have also demonstrated the impact of \gls{HFR} (up to 120 \gls{fps}) to increase the perceived video quality. Moreover, this study shows that \gls{HFR} improves the quality of video content with camera motion, while lower frame-rates are more convenient for sequences with complex motion such as dynamic textures. The study conducted in~\cite{salmon2013higher} by the \gls{BBC} showed that down-conversion of \gls{HFR} video to 50 \gls{fps} would inevitably result in considerable quality degradation. Mackin {\it et al.} in~\cite{mackin2015study} have presented a new database containing videos at different frame-rates from 15 \gls{fps} to 120 \gls{fps}. Subjective evaluations performed on the video database have demonstrated a relationship between the frame-rate and the perceived video quality and confirmed that this relationship is content dependent.

The main limitation of the \gls{HFR} in practical transmission chain is the significant increase in coding and decoding complexities. The complexity increase may prevent the deployment of the \gls{HFR} since the recent SW/HW codecs do not support real-time processing of high resolution video at high frame-rate (+100 \gls{fps}). This complexity overhead, estimated to 40\% of the encoding and decoding times, is required to increase the frame-rate from 60 to 120 \gls{fps}. Moreover, the \gls{HFR} may increase both bit-rate and energy consumption compared to lower frame-rates without significant quality improvements depending on the video content, as shown in~\cite{mackin2015study, mackin2017investigating}.

A number of research works have investigated \gls{VFR}~\cite{ma2012modeling,huang2016perceptual,katsenou2018vfr,afonso2018video}. These \gls{VFR} solutions use different motion-related features with tresholding techniques~\cite{ma2012modeling,afonso2018video} or \gls{ML} algorithms~\cite{huang2016perceptual,katsenou2018vfr} to select the desired frame-rate. The main limitations of these solutions are either a static sequence-level frame-rate adaptation, which greatly reduces the possible coding gains compared to a dynamic adaptation, or the target application with frame-rates lower than \textit{30fps} or \textit{60fps}, making them unusable for the recent \gls{HFR} format without major updates and thorough testing.

In this paper we propose a content-dependent variable frame-rate solution that determines the critical frame-rate of \gls{HFR} videos. The critical frame-rate is the lowest possible frame-rate that does not affect the perceived video quality of the original \gls{HFR} video signal. The proposed solution is based on a machine learning approach that takes as an input spatial and temporal features of the video to determine as an output the critical frame-rate. This can be considered as a classification problem to derive the critical frame-rate among three possible considered frame-rates: 30, 60 and 120 \gls{fps}. The motivation behind these three frame rates is mainly for compliance with the frame rates specified by ATSC~\cite{ATSC} and DVB~\cite{DVB} for use with various broadcast standards. The subjective results conducted on ten \gls{HFR} video contents clearly show the efficiency of the proposed solution enabling to determine the lowest possible frame-rate while preserving the quality of the \gls{HFR} content. This \gls{VFR} solution enables significant bit-rate savings and complexity reductions at both encoder and decoder.

The rest of this paper is organized as follows. Section~\ref{sec:related-work} gives a short overview of related works on \gls{HFR} video, including coding, quality evaluation and rendering, followed by the objective and motivation of the paper. The proposed variable frame-rate solution is investigated in Section~\ref{sec:rf-classif} as a classification problem with two binary \gls{RF} classifiers. The ground truth generation and features extraction, used to train the two \gls{RF} classifiers, are described in Section~\ref{sec:ground-truth}. Section~\ref{sec:training} gives details on the training of the two \gls{RF} classifiers. The performance of the proposed variable frame-rate solution is assessed in Section~\ref{sec:results} in terms of perceived video quality, compression and complexity efficiencies. Finally, Section~\ref{sec:conclusion} concludes the paper.   
  
\section{Related Work}
\label{sec:related-work}

\subsection{High Frame-Rate Video}
\label{subsec:hfr-base}

The \gls{UHDTV} signal, defined in the \gls{ITU}-R BT. 2020 recommendation~\cite{itu-r_recommendation_BT2020-1}, introduces a number of improvements over the \gls{HDTV}~\cite{itu-r_recommendation_BT709} aiming at providing a better visual experience to the user. Along with a wider color gamut and an increased bitdepth, which allow to depict real colors and avoid ringing artifacts respectively, the key features of the \gls{UHDTV} signal enabling a better depiction of live content are the higher spatial resolution - up to 3840x2160 and 7680x4320 pixels - and increased frame-rate - up to 120 \gls{fps}. The different experiments that lead to the definition of each characteristic of the \gls{UHDTV} signal are summarized in~\cite{nilsson2015ultra,rep2246}.

Particularly, high frame-rate video has been an active field of research in the last decade, with the goal of avoiding well-known motion-related artifacts, namely flickering, motion blur and stroboscopic effect, which are present in traditional \gls{HDTV} frame-rates of 60 \gls{fps} and lower. Flicker is a phenomenon in which unwanted visible fluctuations of luminance appear on a large part of the screen and occurs at low refresh rates on non hold-type displays (e.g. \glspl{CRT}). Several studies~\cite{UHDTV_IEEE,emoto2012critical} have shown that flicker can be eliminated, for \gls{UHDTV} signals, by simply using a frame-rate higher than 80 \gls{fps}. The stroboscopic effect is the result of temporal aliasing, where the frame-rate is insufficient to represent smooth motion of objects in a scene causing them to judder or appear multiple times. At a given frame-rate, strobing can be reduced by lowering the shutter speed of the camera. However, a lower shutter speed also increases motion blur, which is caused by the camera integration of an object position over time, while the shutter is opened. Thus, strobing artifacts and motion blur can not be optimized independently except by using a higher frame-rate~\cite{salmon2011higher}. 

Based on previous studies by Barten~\cite{barten1999contrast} and Daly~\cite{daly2001engineering}, Laird~\textit{et al.}~\cite{laird2006spatio} defined a spatio-velocity \gls{CSF} model of the \gls{HVS} taking into account the effect of eye velocity on sensitivity to motion. In~\cite{noland2014application}, Noland uses this model along with traditional sampling theory to demonstrate that the theoretical frame-rate required to eliminate motion-blur without any strobing effect is 140 \gls{fps} for untracked motion and as high as 700 \gls{fps} if eye movements are taken into account. Since this theoretical critical frame-rate is not yet achievable, several subjective studies have investigated the frame-rate for which motion-related artifacts are acceptable for the \gls{HVS}. In~\cite{selfridge2016visibility}, Selfridge~\textit{et al.} investigate the visibility of motion blur and strobing artifacts at various shutter angles and motion speeds for a frame-rate at 100 \gls{fps}. Their subjective tests showed that even at such a frame-rate, all motion-blur and strobing artifacts can not be both avoided simultaneously. Kuroki~\textit{et al.}~\cite{kuroki2007psychophysical} conducted a subjective test with frame-rates ranging from 60 to 480 \gls{fps}, concluding that no further improvements of the visibility of blurring and strobing artifacts were visible above 250 \gls{fps}. Recently, Mackin~\textit{et al.}~\cite{mackin2017high} have performed subjective tests on the visibility of motion artifacts for frame-rates up to 2000 \gls{fps}, achieved using a strobe light with controllable flash frequency. The study concluded that a minimum of 100 \gls{fps} was required to reach tolerable motion artifacts.

For the purpose of the \gls{UHDTV} signal definition, several studies further investigated the importance of \gls{HFR} for television~\cite{emoto2014high,salmon2013higher,hulusic2017quality}. Emoto~\textit{et al.}~\cite{emoto2014high} showed that increasing the frame-rate from the traditional 60 \gls{fps} to high frame-rate of 120 \gls{fps} provides a significant visual quality improvement. It is also stated that a further increase to 240 \gls{fps} would also improve the motion portrayal but to a much lesser extent than the transition from 60 to 120 \gls{fps}. Salmon~\textit{et al.}~\cite{salmon2013higher} have also studied \gls{HFR} for television, showing that at least 100~\gls{fps} is required for improvements over \gls{HDTV}, especially for content with high motion such as sports. Recently, with one of the first 65 inches \gls{UHD} \gls{HFR} prototype displays, Hulusic~\textit{et al.}~\cite{hulusic2017quality} studied the joint and independent contributions of 4K resolution and \gls{HFR}. The subjective tests carried out showed that the 2160p100Hz format enables a significant increase in visual quality over other configurations - 1080p50Hz, 1080p100Hz and 2160p50Hz - but also that the improvements are strongly content dependent. 

\subsection{Compression of HFR content and Variable Frame-Rate}
\label{subsec:hfr-coding}

Since the adoption of \gls{HFR} in the future television standard, through the second phase of the \gls{DVB} \gls{UHD} standard~\cite{EBU-DVB-UHD}, several studies of \gls{HFR} content compression have been carried out. Authors in~\cite{mackin2017investigating} investigated the impact of high frame-rate on video compression, focusing on the perceptual quality of different motion types and frame-rates at several bit-rates. Using the test sequences of the public \gls{HFR} dataset described in~\cite{mackin2015study} compressed using an \gls{HEVC} encoder, it is shown that \gls{HFR} is beneficial and desirable, especially at high bit-rates and even at the current \gls{HDTV} broadcast data rates for sequences containing camera and/or simple motion, for which the encoder can make use of the increased temporal correlation to predict adjacent frames. Sugito~\textit{et al.}~\cite{sugito2018hfr8k} showed that the overhead, in terms of bit-rate, introduced by the increase from 60 to 120 \gls{fps}, is reasonable, with an optimal bit allocation of $6$-$7\%$ of the total bit-rate for the additional frames needed to achieve \gls{HFR} capability. However, one of the main limitation of doubling the frame-rate is the additional encoding complexity, with a near $40\%$ increase in encoding time. 

\gls{VFR}, where the image frequency can be adapted based on the signal characteristics, is one of the solutions to cope with the complexity and bit-rate increases. Authors in \cite{ma2012modeling} have studied the impact of both frame-rate and \gls{QP} on the perceived video quality.  They also developed an accurate rate model and quality model based on single layer and scalable video bitstreams. These models have been applied to frame-rate based adaptive rate control for single-layer and scalable video encodings. However, these models have been designed for frame-rate decisions between values ranging from \textit{7.5fps} to \textit{30fps}, which corresponds to a very low motion portrayal quality compared to \textit{120fps} videos. In \cite{huang2016perceptual}, a \gls{SVR} is used to predict a satisfied user ratio - percentage of people who do not see the difference between original and lower frame-rates - which is then used to dynamically select the appropriate image frequency at a \gls{GOP} or sequence level. The trained \gls{SVR} uses complex and computationally demanding features, notably a visual saliency map and a spatial randomness map for each frame, thus making it unsuitable for real-time dynamic frame-rate selection. In addition, the training set is only composed of up to 60 \gls{fps} content, limiting the frame-rate choice to \textit{60fps}, \textit{30fps} and \textit{15fps}. With their design targeting lower than \textit{120fps} maximum frame-rates, there is no guarantee that the features used by the \gls{VFR} models proposed in~\cite{ma2012modeling} and~\cite{huang2016perceptual} would also work on \textit{120fps} content due to the different motion portrayal observed in \gls{HFR} content. 

More recently, \gls{VFR} for \gls{HFR} content has been investigated, aiming at offering a perceptually indistinguishable temporally downsampled video~\cite{katsenou2018vfr,afonso2018video}. Katsenou~\textit{et al.} train Bagged Decision Trees to predict the critical frame-rate at a sequence level~\cite{katsenou2018vfr}. The selected feature set is only composed of an \gls{OF} for the  temporal aspect, and \gls{GLCM} for the spatial details contribution. In addition, the considered dataset consists of 22 test sequences with critical frame-rates of 60 or 120 \gls{fps}. Thus, a good generalization of the \gls{VFR} decision problem is hard to achieve with such few data to train and validate the model. In~\cite{afonso2018video}, a dynamic frame-rate adaptation is proposed based on the frame-rate dependent metric FRQM~\cite{zhang2017frame}. The temporal adaptation is coupled to a spatial resolution adaptation using kernel-based downsampling and a neural network based upsampling at respectively the pre and post-processing stages. The spatio-temporal adaptation model shows high coding gains through both objective and subjective tests. However, the authors indicates that the temporal adaptation has only been used for one sequence in their dataset, which only contained \textit{60fps} sequences, making the performance evaluation of the \gls{VFR} part of the solution difficult, especially for \gls{HFR} content.

\subsection{Motion Blur Rendering and Video Frame Interpolation}
\label{subsec:frame-decim-interp}

In a pipeline using a \gls{VFR} video format to transport the video, several processing steps could be added to improve the perceptual quality of the output video. Indeed, on one hand, motion blur can be synthesized when the frame-rate is lowered to render a video close to what would have been captured with a camera at the lower frame-rate and its corresponding shutter speed. This would reduce the stroboscopic effect due to the frame decimation thus improving the visual quality of the \gls{VFR} video. Motion blur synthesis has been extensively studied in an effort to render synthetic images as real as possible~\cite{navarro2011motion}. These techniques mostly rely on the perfect knowledge of the depth and motion of the scene and are thus not compatible for a live broadcast use-case. More recently, a motion blur rendering algorithm using only two consecutive images as inputs to produce a motion blurred output have been designed in~\cite{brooks2019learning}. These promising results are balanced by the computationally demanding algorithm, due to the underlying \gls{CNN} architecture used to synthesize motion blur.

\begin{figure*}[t]
\centering
\includegraphics[width=0.7\textwidth]{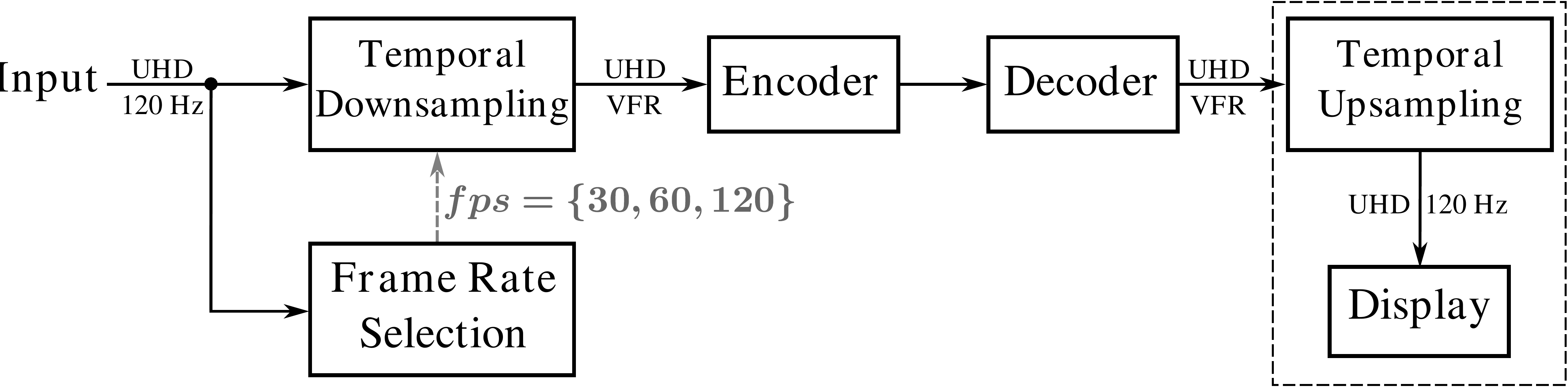}
\caption{Block diagram of the complete Variable Frame-Rate (VFR) coding scheme.}
\label{fig:global-scheme}
\end{figure*}

On the other hand, since most display devices do not support a variable frame-rate, the \gls{VFR} video must be temporally upsampled to the original higher frame-rate before displaying it. Thus, frame interpolation methods can be used to improve the temporal upsampling step in order to obtain a visually better displayed video. Video frame interpolation is a well-studied field with several existing approaches to the problem. The classical approach interpolates intermediate frames from the optical flow field~\cite{baker2011database} of the scene. Interpolated frames, whose quality highly depends on the accuracy of the computationally expensive optical flow computation~\cite{sun2014quantitative}, typically suffer from motion boundaries and severe occlusions thus showing strong artifacts, even with state-of-the-art optical flow algorithms~\cite{ilg2017flownet}. More recent promising works rely on neural networks to either predict convolution kernels for each pixel used to generate the interpolated frames~\cite{niklaus2017video} or leverage optical flow fields with exceptional motion maps~\cite{park2020robust}. However, these techniques involve a large number of convolutions, sometimes with large kernels (up to 41x41 for each pixel) to cope with large motion, thus making the computational demand unsuitable for real-rime use-cases.

\subsection{Objective and Motivation}
\label{subsec:obj-and-motiv}

Most existing algorithms containing variable frame-rate have been designed purely for rate control in 30 \gls{fps} video encoding schemes, with the goal of skipping frames when the bit budget constraint can not be met. Such behavior does not take properly into account the impact on perceptual quality, making these solutions not suitable for \gls{HFR}, which has been integrated to the \gls{UHDTV} standard to improve motion portrayal. 

Since the improvements brought by \gls{HFR} capabilities are highly content dependent, several recent studies have based the frame-rate selection on perceptual factors to lower the frame-rate when there is no impact on visual quality. However, they use computationally expensive features and rely on a small dataset, not always composed of 120 \gls{fps} content, to train and validate the variable frame-rate models. Thus, these solutions do not achieve good generalization of the problem. Moreover, they are not suitable for a real-time constraint, which is required for use-cases like live broadcast of, for example, sport events, that would periodically highly benefit from \gls{HFR}.

In this paper, real-time variable frame-rate for \gls{HFR} source content is addressed. The proposed system, depicted in~\Figure{\ref{fig:global-scheme}}, relies on two classifiers to predict the critical frame-rate. It has been designed with the following objectives:
\begin{itemize}
\item Design a dynamic frame-rate selection at a \gls{GOP} level with perceptually invisible frame-rate changes.
\item {Low-complexity feature computation with low impact on the coding chain processing time}.
\item Training and testing on a well dimensioned dataset containing various types of 120 \gls{fps} content.
\item Perceptual validation of the obtained objective performance through subjective evaluation test.
\item Assess the bit-rate and complexity savings within an \gls{HEVC} encoding chain.
\end{itemize} 

To meet the real-time constraint of live broadcast, state-of-the-art motion blur rendering and frame interpolation have not been integrated in the \gls{VFR} pipeline. Instead, the proposed system has been designed with simple frame decimation (resp. duplication) as a temporal downsampling (resp. upsampling) tool. {Additionaly, the \gls{RF} algorithm has been chosen as classification technique thanks to a small benchmark comparing several \gls{ML} techniques for \gls{VFR} classification that showed a better trade-off between prediction accuracy and computational complexity for the \gls{RF} algorithm.} 

\section{Random Forest Classifier for Variable Frame-Rate}
\label{sec:rf-classif}

This section briefly presents Random Forests (RFs) as a classification tool and introduces the method proposed in this work to reduce the frame-rate with no visual impact, i.e. the \gls{VFR} decision problem, as a combination of two binary classification problems.

\subsection{Background on Random Forests}
\label{subsec:rf-classif-protocol}

Random Forests~\cite{breiman2001random} are a common \gls{ML} tool used to solve classification problems. A \gls{RF} classifier is able to predict the value of a target variable, i.e. a class, based on a set of input variables, i.e. input features, using the majority vote of an ensemble of nearly independent decision trees.

A decision tree is constructed by first partitioning the training dataset, i.e. the features and the associated class of each sample, into two different subsets, called nodes. This process is performed recursively until either all the node samples belong to a single class or a tree constraint has been reached. At each node, each available input feature is evaluated for all its possible values, in order to achieve the best separation of the classes in the subsequent child-nodes. 

In this work, the criterion used to quantify the quality of a split, given the feature $F$ and its threshold value $t$, is based on the Gini impurity measure, a common metric for Decision Trees~\cite{breiman1984classification}. It is computed as follows
\begin{equation}
  I_{G}(D) = \sum_{c\, \in \, C} \ensuremath{\mathbb{P}}(c\, |\, D) \, \left(1-\ensuremath{\mathbb{P}}(c\, |\, D)\right),
  \label{eq:gini}
\end{equation}
with $D$ the sample set under consideration, $C$ the set of possible class labels and $\mathbb{P}(c\, |\, D)$ the conditional probability of class $c$ given the sample set $D$. 

The best split is then obtained by finding the pair $(F,t)$ that maximizes the \gls{MDI} $\Delta I_{G}$ defined by Equation~(\ref{eq:mdi}).
\begin{equation}
  \Delta I_{G}(D,F,t) = I_{G}(D) - \dfrac{|D_{L}|}{|D|} \, I_{G}(D_{L}) - \dfrac{|D_{R}|}{|D|} \, I_{G}(D_{R}), 
  \label{eq:mdi}
\end{equation}
with $D_{L} = \{x\in D, F(x)<t\}$ (resp. $D_{R} = \{x\in D, F(x)\geq\}$) the subset of sample set $D$ for which each sample $x$ has a value of feature $F(x)$ smaller (resp. larger) than threshold $t$ and $|D|$ the cardinal of a set $D$.

To minimize the correlation between trees of the \gls{RF}, the bootstrap aggregating, or bagging, technique~\cite{breiman1996bagging} is used to construct the forest. This consists in training each tree $T_{i}$ with a different subset $D_{i}$ of the input data sample set $D$. Each $D_{i}$ is obtained by a uniform sampling of $D$ with replacement, i.e. replacing discarded samples by duplicates of a selected one. In addition and to further reduce the correlation between trees, only a random subset of the features, here $\sqrt{n}$ features with $n$ the total number of input features, are evaluated at each node to find the best available split.

\subsection{VFR Classification Problem}
\label{subsec:rf-classif-composition}

The proposed solution aims at predicting when the frame-rate can be reduced, by discarding frames, without any perceptual impact on the quality of the original input \gls{HFR} video. In an effort to keep the number of possible frame-rates reduced and obtain a regular frame decimation process, two frame-rates, \textit{60 fps} and \textit{30 fps}, were identified as potential candidates in addition to the original frame-rate of \textit{120 fps}. The \gls{VFR} decision problem thus becomes a three-class classification. 

\begin{figure}[!t]
\centering
\includegraphics[width=0.8\linewidth]{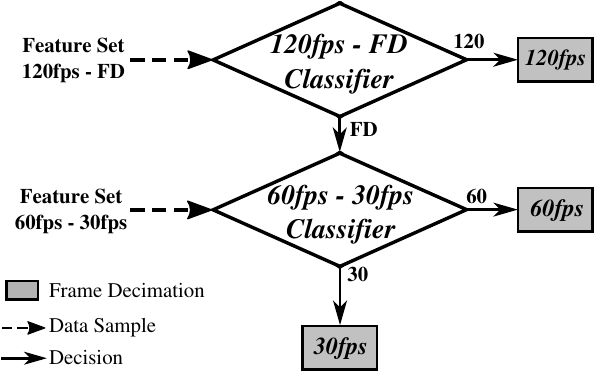}
\caption{Overall prediction scheme with cascaded binary RF classifiers.}
\label{fig:casc-scheme}
\end{figure}

In this work, a combination of two successive binary \gls{RF} classifiers has been chosen to solve the classification problem, as depicted in \Figure{\ref{fig:casc-scheme}}, instead of directly training a forest with multi-class outputs. This decision leads to a better overall performance by training both classifiers independently on separate datasets and features. Indeed, in addition to the specialization of each binary classifier, almost all samples of the database can be used for training either one or both classifiers while keeping balanced training datasets, as described in Section~\ref{sec:ground-truth}, thus increasing the accuracy of the overall model. 

The first \gls{RF} classifier, named \textit{120fps - FD}, is specialized in deciding whether the frame-rate must remain \textit{120 fps} or a \gls{FD} can be applied without impacting the visual quality. If the \textit{120 fps} class is chosen by the first classifier, the frame-rate prediction process is stopped and no \gls{FD} is applied. Otherwise, the prediction process continues by requesting the second classifier, named \textit{60fps - 30fps}, which aims at selecting the appropriate lower frame-rate if a \gls{FD} is applied on the input \gls{HFR} video. 

\section{Ground Truth Generation}
\label{sec:ground-truth}

One of the most crucial steps towards training a supervised \gls{RF} model is to gather a dataset which will be used as ground truth, i.e. examples - features representing a sample and the sample class label - used by the model to learn to predict. It is thus important to have a ground truth that contains a good representation of all the real cases the model could encounter in order to achieve a good generalization of the problem. This section focuses on detailing the ground truth generation process, necessary to obtain the datasets used to train both \gls{RF} classifiers. The \gls{HFR} database is first presented, followed by the detailed methodology for subjectively determining the critical frame-rate labels. Then, the creation of the dataset is described, with the composition of the balanced training sets on one hand and the feature extraction process on the other hand.

\subsection{HFR Video Database}
\label{subsec-hfr-database}

To the best of our knowledge, there is no publicly available dataset for the \gls{VFR} classification problem except for the \gls{BVI-HFR} database~\cite{mackin2015study}. However, this database only contains 22 videos, which is rather small to train a reliable model. In addition, the temporal downsampling technique used to create the lower frame-rates in this database is frame averaging, whereas the chosen method in this work is frame decimation, which could lead to different decisions on the critical frame-rate.

Therefore, a new database has been gathered, composed of 375 native \gls{HFR} video clips of 5 to 10 seconds. These clips are all uncompressed and stored in YUV format with 4:2:0 chroma subsampling and 8-bit depth. Their original frame-rate is 120 fps and spatial resolution {1920$\times$1080} pixels (\gls{HD}) - a downsampling has been performed for source videos of higher spatial resolution using Lanczos3 filter banks~\cite{duchon1979lanczos}.

The database includes video clips from the BVI-HFR dataset~\cite{mackin2015study}, together with b\textless\textgreater com, Harmonic and other non-publicly available test sequences. In order to later evaluate the trained model on unseen data samples, 15 sequences, with heterogeneous spatio-temporal characteristics, have been extracted from the database, leaving 360 video clips to annotate before training both \gls{RF} classifiers. 

\subsection{Critical Frame-rate Decision Methodology}
\label{subsec:ground-truth-background}

The ground truth generation process requires each video of the database to be assigned a critical frame-rate chosen among the three considered ones in the \gls{VFR} classification problem. A subjective test has thus been carried out with the objective of finding the lowest frame-rate for which no visual degradation can be observed compared to the original 120 fps video. Following the \gls{ITU}-R BT.500-13 recommendation~\cite{recBT500}, the \gls{SDSCE} protocol has been used with a binary scale - either identical or visible difference - and two screens placed side-by-side. Therefore, subjects were asked, for each video of the database, if there was a visible difference between the two displayed videos, i.e. the known reference and the lower frame-rate, either 30 or 60 fps, test video.

\begin{figure}
\centering
\includegraphics[width=\linewidth]{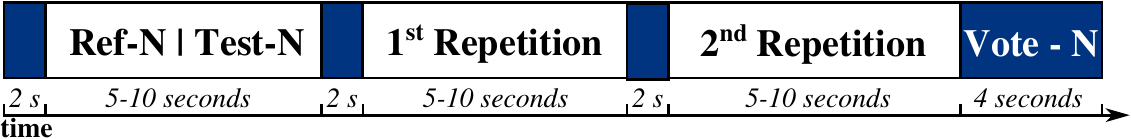}\label{subfig:annot-process-btc}
\caption{Basic Test Cell (BTC) for the SDSCE evaluation method.}
\label{fig:annot-process}
\end{figure}

The test was composed of 750 \glspl{BTC}, randomly divided into 30-minute test sessions. Each \gls{BTC} is 40-second long and is composed of a 2-second message announcing the test video index followed by the side-by-side display of the reference and test videos with two repetitions.

A 2-second break displaying a mid-gray image has been added between each repetition. The \gls{BTC} is concluded by a 4-second message asking the viewer to vote, as shown in~\Figure{\ref{fig:annot-process}}. Due to the large duration of the subjective test, only five viewers, experts in video processing, participated in the whole database annotation. The final frame-rate decision is the lowest frame-rate for which the majority of expert viewers did not notice any visible difference with the reference 120 fps video.

The tests were conducted in a controlled laboratory environment, following the ITU-R Rec BT.500-13~\cite{recBT500}. Two identical 27-inch screens capable of displaying 120 fps content (Asus RoG  Swift PG278Q) were used side-by-side, aligned and placed at a distance from the viewer position of three times the screen height. Each participant has been screened to ensure (corrected to) normal visual acuity and normal color vision.

\subsection{Balanced Dataset Composition}
\label{subsec:ground-truth-problem}

The results of the expert subjective test are summarized in~\Table{\ref{tab:database}}. As can be seen, the sequences are not evenly distributed over the three possible frame-rates due to a large proportion of the available content being captured with \gls{HFR}-capable devices containing high motion content, for which frame decimation downsampling to 30 fps is critical. Since the goal is to allow for a frame-rate adaptation at the lowest possible level, i.e. a frame-rate decision for every chunk of 4 frames to keep a regular frame decimation process, the sequence-level labels obtained via the subjective test have been extended to 4-frame chunk-level labels. Thus, for sequences with significant motion discontinuities, video shots with uniform motion can be identified and assigned different labels in a single sequence. Based on the observations made by the experts after the subjective test, a total of 429 video shots with uniform motion have been extracted from the 360 native \gls{HFR} sequences. This refinement allows for a more accurate annotation of the database, thus avoiding chunks being annotated with inconsistent labels. For instance, a chunk not containing any movement associated to the \textit{120fps} class. However, such cases could still remain in the ground truth due to the difficulty to identify the motion discontinuities with a precision of less than four frames.

\begin{table}[t]
\centering
\normalsize
\caption{Database critical frame-rate distribution.}
\label{tab:database}
\footnotesize
\begin{tabular}{l|c|c|c|}
                              & \multicolumn{3}{c|}{Critical frame-rate} \\[0.5em]
                              &     30      &     60     &    120      \\ \midrule[0.3mm]
\# of shots w/ uniform motion &     78      &     184    &    167      \\[0.3em]
\# of 4-frame chunks          &    3749     &    22327   &   19996     \\ \midrule[0.3mm]
\end{tabular}
\end{table}

From this ground truth, two different datasets were created to train the \textit{120fps-FD} and \textit{60fps-30fps} \gls{RF} classifiers. The first one contains all samples of the \textit{120fps} class as well as those from the \textit{\gls{FD}} class. The \textit{\gls{FD}} class is composed of all \textit{30fps} samples and a random subset of the \textit{60fps} samples. The amount of selected \textit{60fps} samples has been chosen to produce a balanced dataset, i.e. to roughly obtain the same sample size for both the \textit{120fps} and \textit{FD} classes. The second dataset, used to train the \textit{60fps-30fps} classifier, is comprised of the \textit{30fps} samples and a random subset of the \textit{60fps} samples, whose size has also been chosen to produce a balanced dataset. The choice of balancing datasets has been made because the unbalanced class distribution in the database does not necessarily represent the distribution of media content in a broadcast context, the chosen use-case for the proposed solution, but rather relates to the current difficulty to find \gls{HFR} content with low motion. This is due to the fact that most of the currently available \gls{HFR} content has been shot to demonstrate the gain in perceptual quality and motion portrayal brought by the technology.

\subsection{Feature Extraction}
\label{subsec:feature-extraction}

The goal of a feature set is to gather the different metrics relevant to the considered classification problem that would help discriminate the output classes from one another. For the \gls{VFR} classification problem, a first feature would intuitively be the motion information, e.g. the motion vectors between two consecutive frames. Indeed, high movement in a source \gls{HFR} video will likely lead to visible temporal aliasing, i.e. stroboscopic effect, if a lower frame-rate is used after frame decimation. In addition, since motion blur is not added during the temporal downsampling process used in this work, lowering the frame-rate could introduce visible jerkiness in high motion videos.

For the three considered frame-rate classes, flickering, the other well-known motion-related artifact, can appear in highly textured areas where the local variation in luminance between two consecutively displayed frames would be visible at lower frame-rates. In an effort to capture this phenomenon in the feature set, the pixel luminance values and directional gradients can be used.

Based on the performed expert viewing sessions, it has been observed that small objects with high velocity, which would not necessarily be detected by the motion vectors depending on the used motion estimation algorithm, could induce visible artifacts at lower frame-rates. To take this observation into account, a simple metric capable of detecting both global displacements and small moving objects has been designed. This metric is based on the thresholding of the difference between two consecutive frames. First, the frame difference $D_{n}(i,j)$, i.e. the difference in pixel value of the luminance plane at the same location in space $(i,j)$ between the $n^{th}$ frame $F_{n}$ and the preceding one $F_{n-1}$, is computed for each pixel using Equation~(\ref{eq:motion-diff})
\begin{equation}
D_{n}(i,j) =  |\,F_{n}(i,j) - F_{n-1}(i,j)\,|.
\label{eq:motion-diff}
\end{equation}

\begin{figure}[t]
\centering
\subfloat[]{\includegraphics[width=0.49\linewidth]{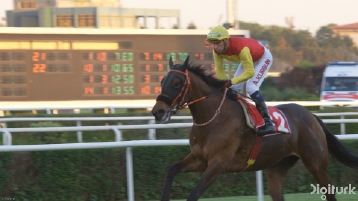}\label{subfig:motion-diff-source}}
\,
\subfloat[]{\includegraphics[width=0.49\linewidth]{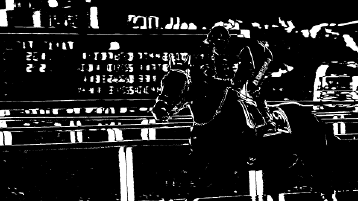}\label{subfig:motion-diff-thresh}}
\caption{Example of thresholded motion difference with (a) the original image of the \textit{Jokey} sequence and (b) the thresholding activation map with threshold $Th = 25$.}
\label{fig:motion-diff}
\end{figure}

Then, a thresholding operation is performed on the frame difference image, defined as follows
\begin{equation}
A_{n,Th}(i,j) = \left\{ \begin{array}{ll} 1  & if\ D_{n}(i,j) \geq Th \\ 0  & if\ D_{n}(i,j) < Th \\\end{array} \right.,
\label{eq:thresh-motion-diff}
\end{equation}
with $A_{n,Th}$ the resulting thresholding activation map for the $n^{th}$ frame and a threshold $Th$. \Figure{\ref{fig:motion-diff}} depicts an example with both the original image and the resulting thresholded frame difference image.

The designed feature set is thus based on the following feature maps:
\begin{itemize}
\item \textbf{NormMV}, \textbf{HorMV}, \textbf{VerMV:} maps respectively representing the \gls{MVs} norm, horizontal coordinate and vertical coordinate.
\item \textbf{ThreshDiffMap:} thresholded frame difference map as defined in Equation~(\ref{eq:thresh-motion-diff}).
\end{itemize}

\begin{itemize}
\item \textbf{GradMag}, \textbf{GradHor}, \textbf{GradVer:} maps respectively representing the Sobel gradient magnitude, horizontal gradient and vertical gradient.
\item \textbf{Luma:} pixel luminance map.
\end{itemize}

For each map, several scores have been computed, namely the mean value, the standard deviation, the maximum value and the mean of the 10\% highest values, to produce a total of 32 different features that will serve as an initial feature set for the training of both the considered \gls{RF} models.

\section{Random Forest Training Process}
\label{sec:training}

Once the ground truth is available and the features computed, the \gls{RF} models can be trained to solve the \gls{VFR} classification problem. This section focuses first on the performance evaluation process, necessary to assess and optimize the quality of the model critical frame-rate prediction task. Then, a feature selection process, used to reduce the initial feature set to only the relevant features for each binary classifier, is presented. Finally, the classification results are presented and analyzed.

\subsection{Model Evaluation Process}
\label{subsec:training-evaluation}

In order to optimize a \gls{ML} classifier, it is necessary to use a metric capable of evaluating the model classification performance to find the best parameters. To do so, several common metrics, namely precision, recall and F1-score~\cite{chinchor1992muc}, can be used. First, the trained model confusion matrix has to be computed from the true and predicted labels of the dataset samples. Then, once the different quantities of the confusion matrix, namely \gls{TP}, \gls{FP}, \gls{FN} and \gls{TN}, are available either as number of samples or normalized probabilities, the precision, recall and F1-score can be computed as follows, with $C = \{c_{1},c_{2}\}$ the set of classes for the binary classifier under test
\begin{equation}
precision(C) = \frac{1}{|C|} \,  \sum_{c_{i}\in C}\frac{TP(c_{i})}{TP(c_{i})+FP(c_{i})},
\end{equation}
\begin{equation}
recall(C)\quad = \frac{1}{|C|} \,  \sum_{c_{i}\in C}\frac{TP(c_{i})}{TP(c_{i})+FN(c_{i})},
\end{equation}
\begin{equation}
F_{1}\textnormal{-}score(C)\; = \frac{2}{|C|} \, \sum_{c_{i}\in C}\frac{precision(c_{i}) \, recall(c_{i})}{precision(c_{i}) + recall(c_{i})},
\end{equation}

As for any binary classification problem, the goal is to maximize the confusion matrix main diagonal values, i.e. the number of \gls{TP} and \gls{TN} representing the correct predictions. This can be achieved by maximizing precision, recall, or F1-score during the training process, depending on the considered classification problem and the criticality of each error type. For the \gls{VFR} classification problem, the main goal is also to minimize the critical errors - predicted frame-rate lower than the ground truth - which would potentially induce visible temporal artifacts thus greatly reducing the output visual quality. To emphasize these critical errors and avoid them in the final model, another performance evaluation metric $M_{crit}$ has been designed, as a combination of the precision of the lower frame-rate class and the recall of the higher frame-rate class, using Equation~(\ref{eq:crit-err})
\begin{equation}
\hspace{-0.5em}\begin{array}{ll}
M_{crit}(C) &= \frac{1}{|C|} \,  \left[precision(c_{1}) + recall(c_{2}) \right], \\[0.5em]
            &= \frac{1}{|C|} \,  \left[\frac{TP(c_{1})}{TP(c_{1})+FP(c_{1})} + \frac{TP(c_{2})}{TP(c_{2})+FN(c_{2})}\right], \\[0.5em]
            &= \frac{1}{|C|}\,  \left[\frac{TP(c_{1})}{TP(c_{1})+FP(c_{1})} + \frac{TN(c_{1})}{TN(c_{1})+FP(c_{1})}\right], \\
\end{array}
\label{eq:crit-err}
\end{equation}
with $C = \{c_{1},c_{2}\}$ the set of ordered classes - frame-rate of $c_{1}$ lower than the frame-rate of $c_{2}$. This metric has been used together with the F1-score to assess the quality of \gls{RF} models for both the feature selection process and hyper-parameter tuning described in the next sections.

\subsection{Feature Selection}
\label{subsec:feature-selection}

In an effort to limit the model complexity and improve its performance, a dimensionality reduction algorithm has been used on the proposed initial feature set. Indeed, by only selecting the relevant features, thus removing features carrying useless information for the considered classification problem, both the feature computation time and training time are greatly reduced. Additionally, model over-fitting is also decreased when the size of the feature set is reduced due to the reduction of noise in the input data and the elimination of highly correlated features, i.e. features that would carry the same information about the target variable.

In this work, a \gls{RFE} process has been used to reduce the dimension of the initial feature set. It consists in recursively evaluating the model performance on a dataset and a feature in which the least important feature is removed after each iteration. The feature importance is computed in terms of mean decrease in Gini impurity, i.e. the average capacity of a feature to reduce the Gini impurity computed at a given tree node, using Equation~(\ref{eq:gini}). When the feature set size reaches the minimum tested dimension of 2, the feature set leading to the best model performance among all the tested dimensions is selected as the final feature set. 

\begin{figure}[!t]
\centering
\subfloat[120fps-FD RF model]{\includegraphics[width=0.49\linewidth]{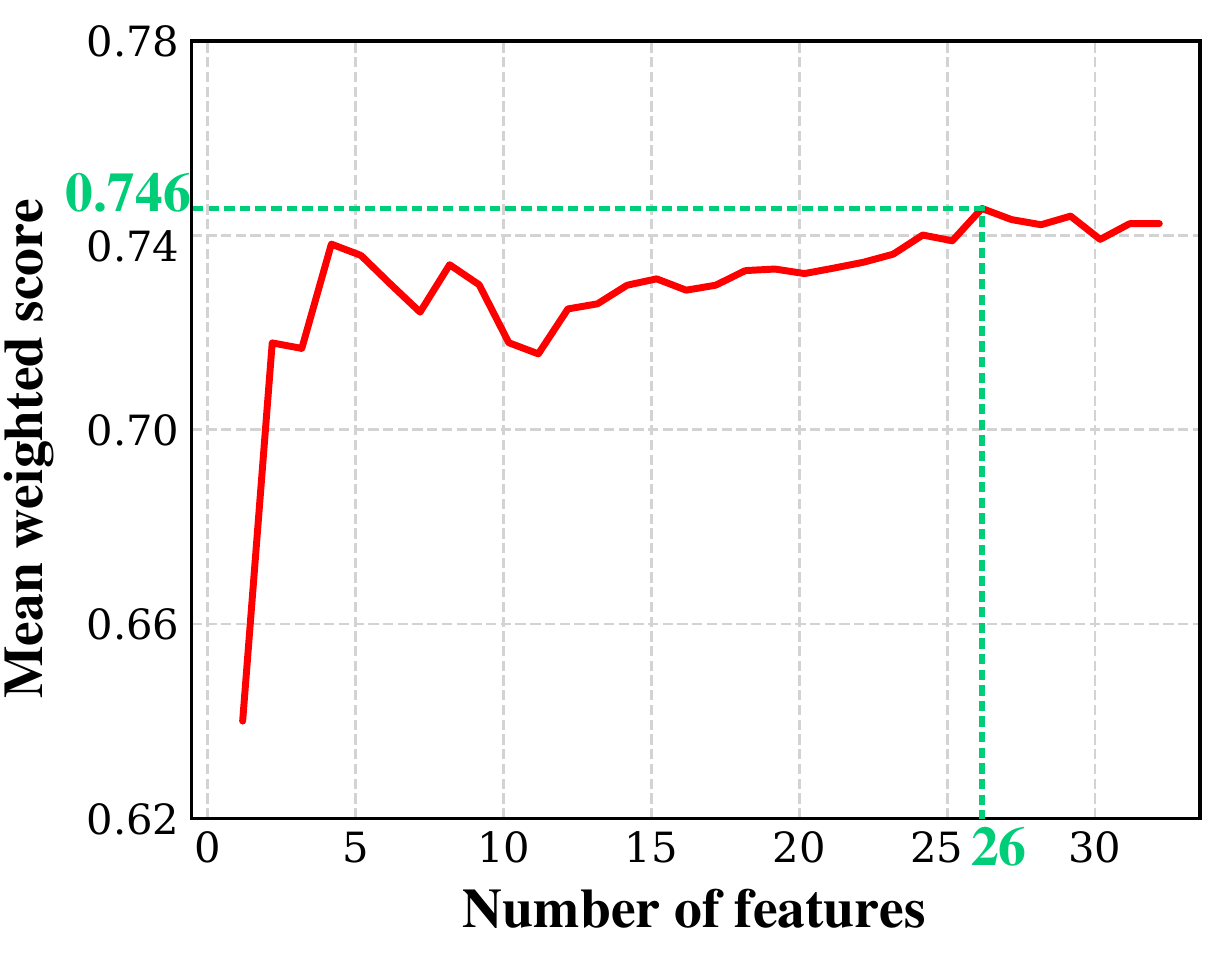}\label{subfig:rfe-rf1}}
\,
\subfloat[60fps-30fps RF model]{\includegraphics[width=0.49\linewidth]{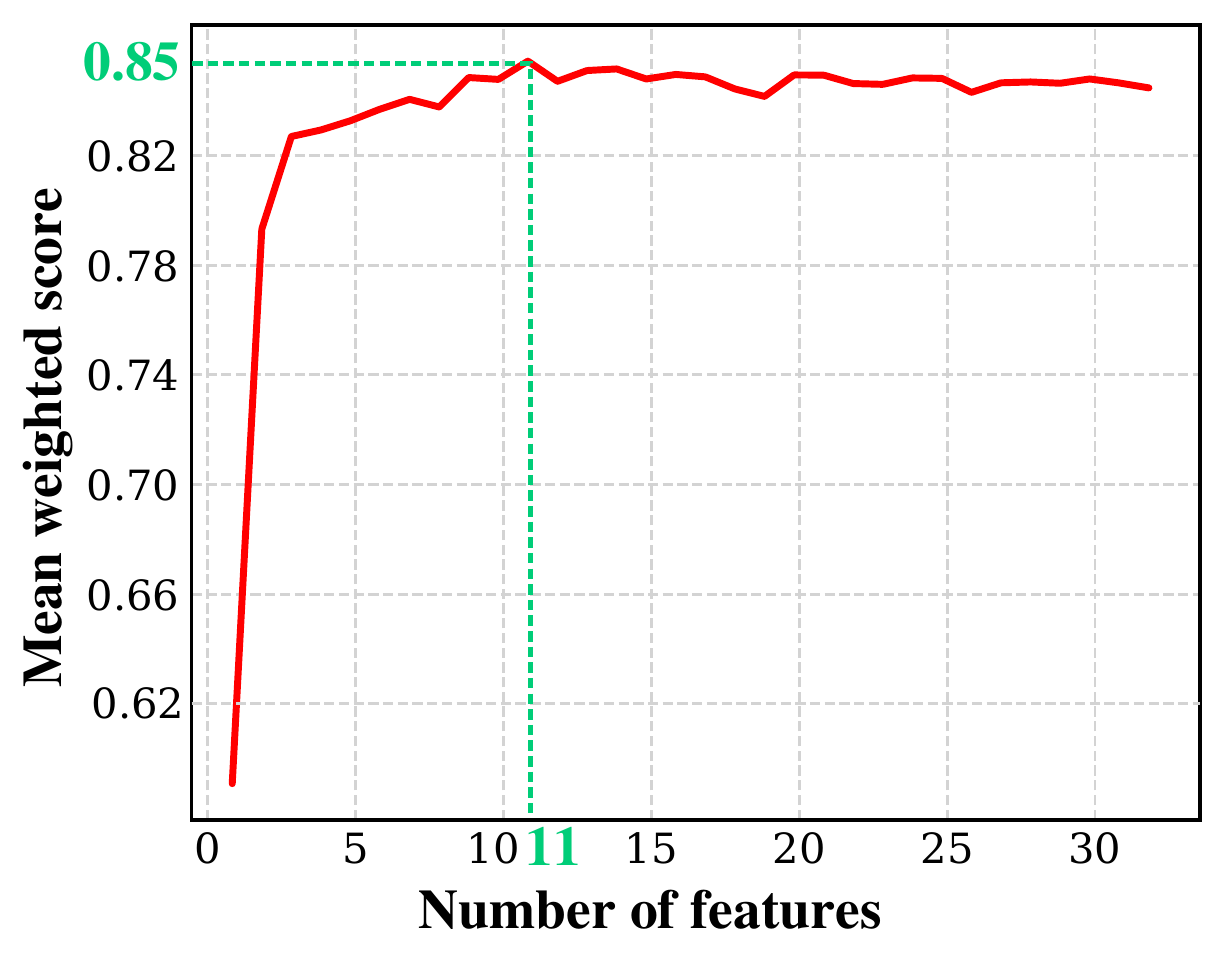}\label{subfig:rfe-rf2}}
\caption{Recursive Feature Elimination with weighted (F1,$M_{crit}$) score.}
\label{fig:rfe}
\end{figure}

\begin{figure*}[!t]
\centering
\subfloat[120fps-FD RF model]{\includegraphics[width=0.65\linewidth]{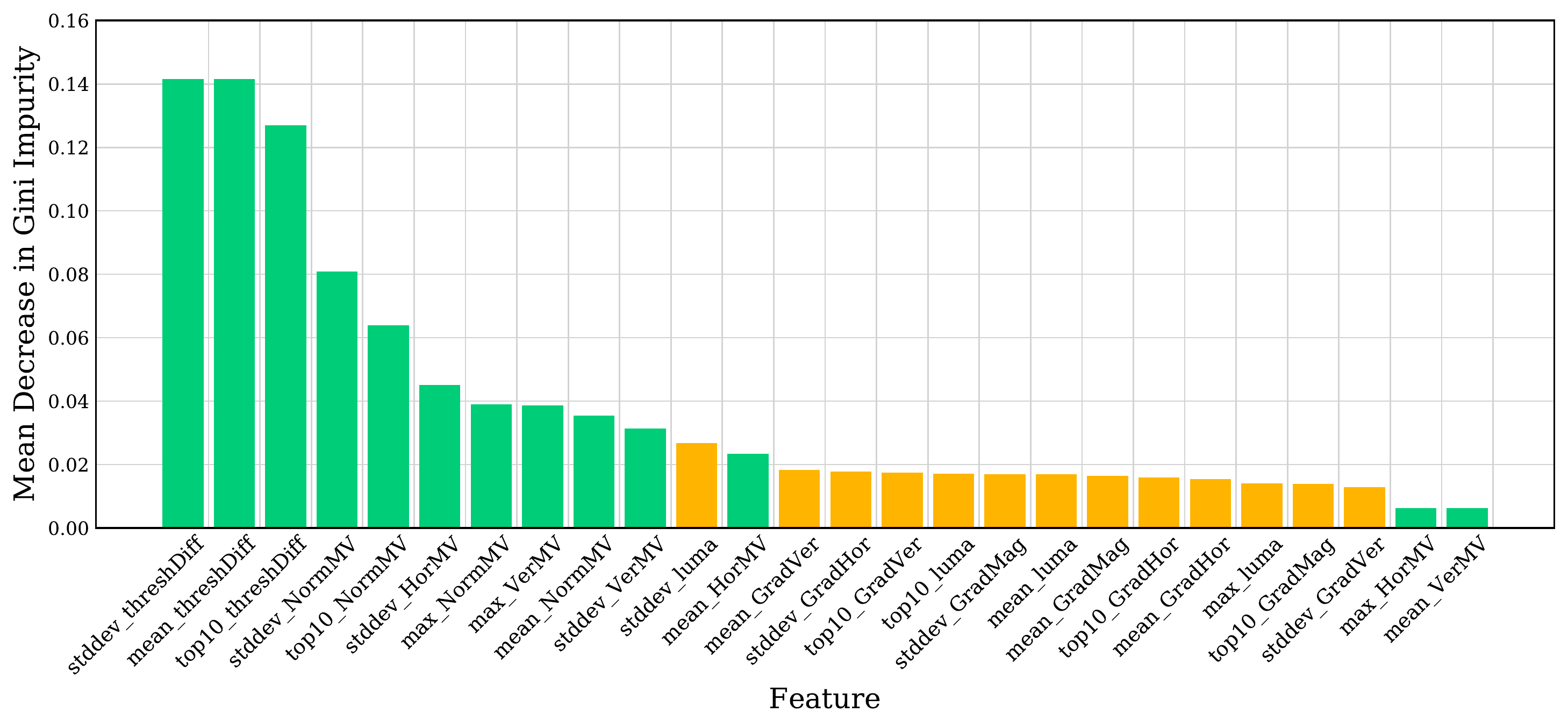}\label{subfig:feature-imp-rf1}}
\,
\subfloat[60fps-30fps RF model]{\includegraphics[width=0.263\linewidth]{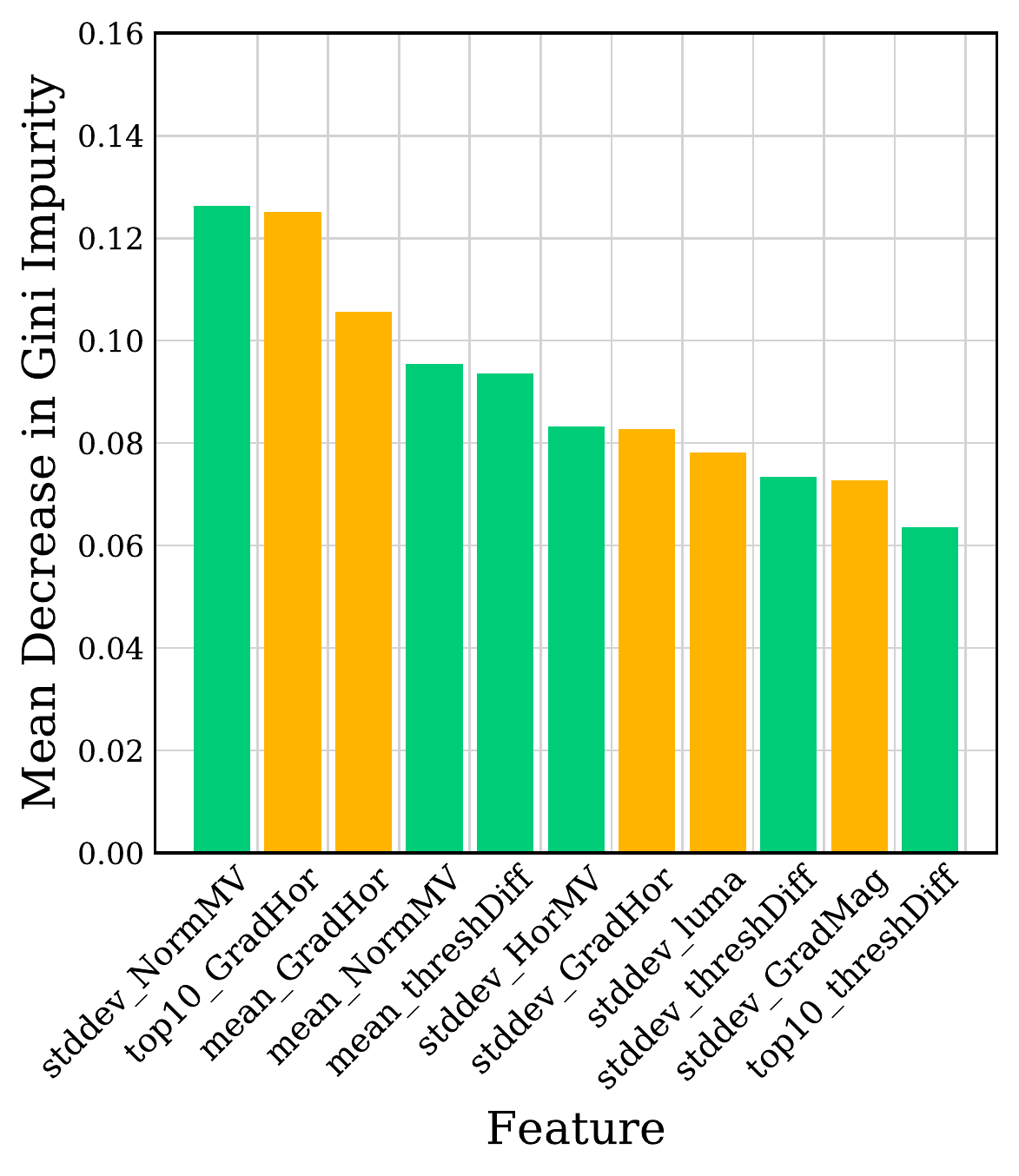}\label{subfig:feature-imp-rf2}}
\caption{Feature importance measured with Mean Decrease in Gini Impurity. \textit{yellow: spatial features, green: motion features.}}
\label{fig:feature-imp}
\end{figure*}

This process has been performed independently for both proposed \gls{RF} models with the same initial feature set but leading to a different optimal feature set size for each binary \gls{RF} classifier, respectively 26 and 11 features for the \textit{120fps-FD} classifier and \textit{60fps-30fps} classifier, as depicted in \Figure{\ref{fig:rfe}}.

\Figure{\ref{subfig:feature-imp-rf1}} shows the list of selected features for the \textit{120fps-FD} \gls{RF} classifier with their corresponding feature importance. As can be observed, the most relevant features to discriminate samples from both classes are based on the two motion features \textit{ThreshDiffMap} and \textit{NormMV}. This correlates well with the observations made by the experts during the ground truth annotation subjective tests. Indeed, it was pointed out that above a certain amount of movement, either from a moving camera or an object with high velocity - which can be captured by both metrics -, a stroboscopic effect as well as jerkiness, due to lack of motion blur, could easily be detected with frame-rates lower than 120 fps. Most of the features based on spatial measures are present in the optimized feature set, with a significantly lower importance compared to the aforementioned motion features. This tends to indicate that flickering becomes an important criterion to keep a high frame-rate when the amount of movement does not induce other motion artifacts.

For the \textit{60fps-30fps} \gls{RF} model, the selected features and their importance are depicted in \Figure{\ref{subfig:feature-imp-rf2}}. As for the first \gls{RF} model, the features based on the motion vectors, \textit{NormMV}, have a high capacity to discriminate samples from both classes. However, spatial features, based on the \textit{Luma} and \textit{GradHor} feature maps, hold a significantly higher importance compared to the first model, indicating that flickering mostly occurs at 30 fps for most of the videos of the training dataset.

\subsection{Classification Results}
\label{subsec:res-classif}

With the final feature sets, an optimization has been conducted on the maximum tree depth and number of trees hyper-parameters, leading to a final model with 200 trees of depth 7 for the \textit{120fps-FD} classifier and 100 trees of depth 7 for the \textit{60fps-30fps} classifier. An in-depth analysis of the prediction capability of the final models, both individually and combined to form the overall \gls{VFR} prediction scheme, can then be conducted. It is important to note that the models have been trained using a 10-fold cross-validation so that the considered performance is a combination of the results from the validation fold of each iteration. This means that each tested sample prediction presented in the different confusion matrices has been obtained without using the validation sample for training. Additionally, the training set has not been shuffled, so that chunks from a same sequence could not be in the training and validation folds at the same time, thus avoiding a highly biased performance evaluation.

\begin{figure}[!t]
\centering
\subfloat[120fps-FD classifier]{\includegraphics[width=0.32\linewidth]{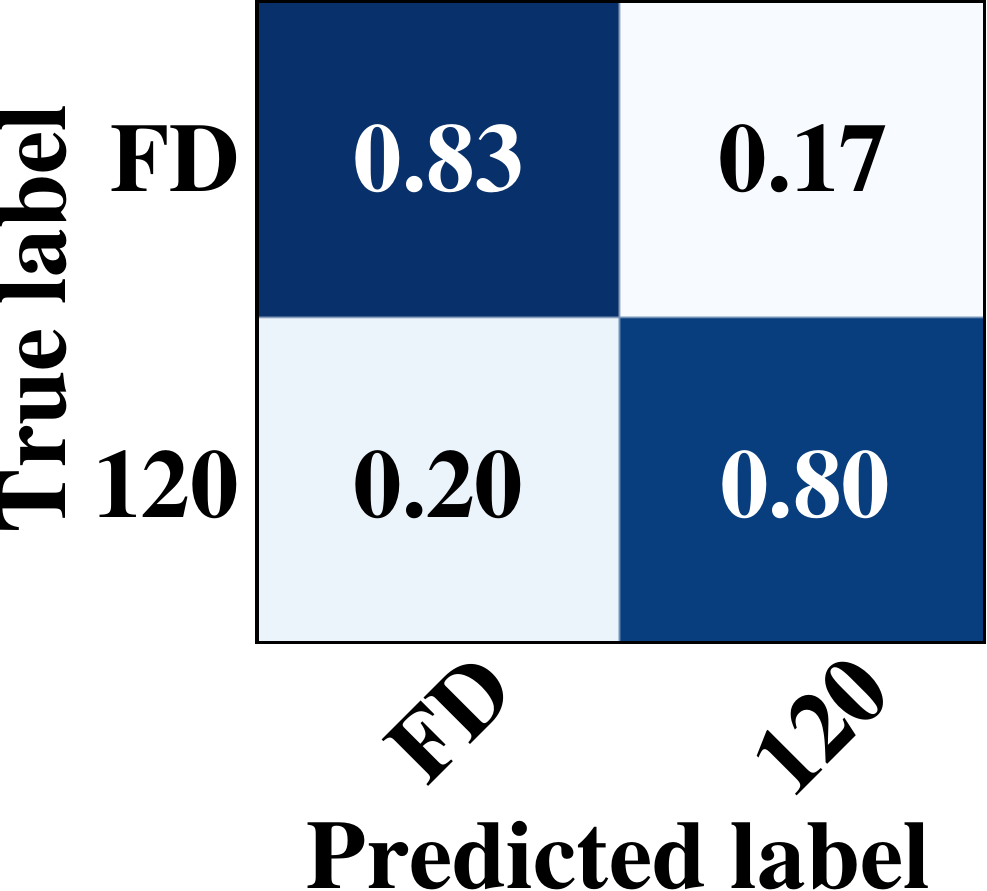}\label{subfig:cm-rf1}}
\qquad\qquad
\subfloat[60fps-30fps classifier]{\includegraphics[width=0.32\linewidth]{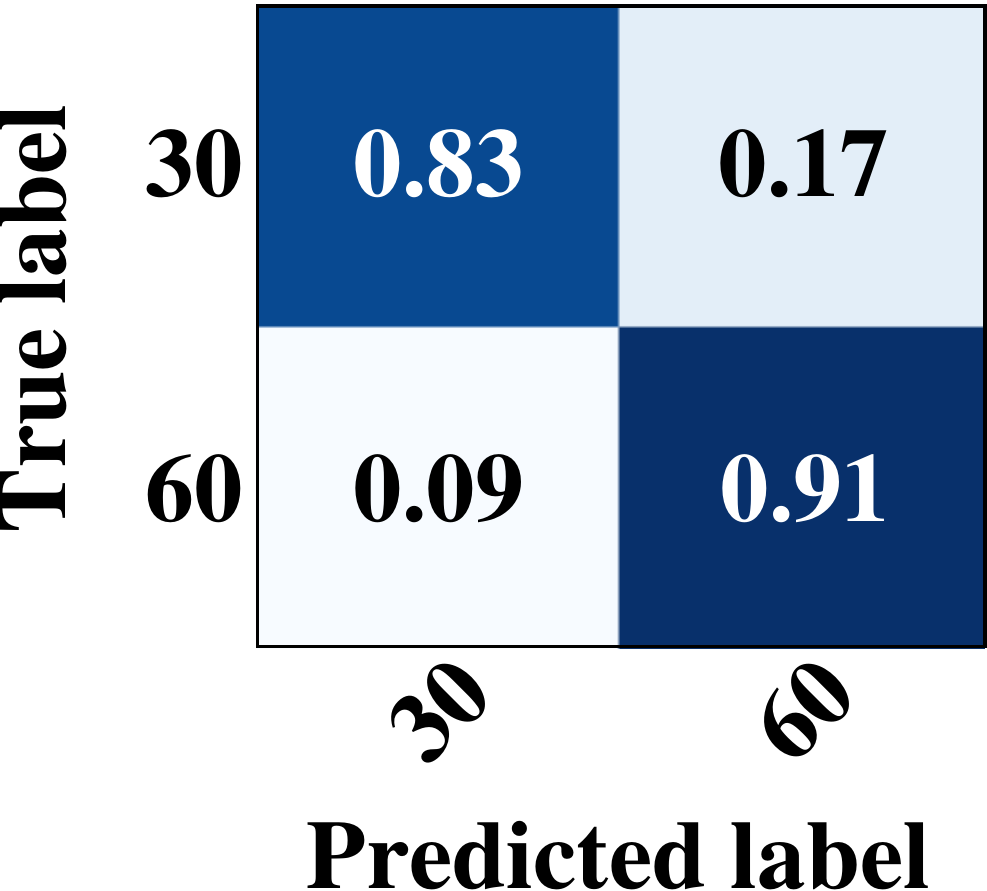}\label{subfig:cm-rf2}}
\\
\subfloat[Overall scheme]{\includegraphics[width=0.38\linewidth]{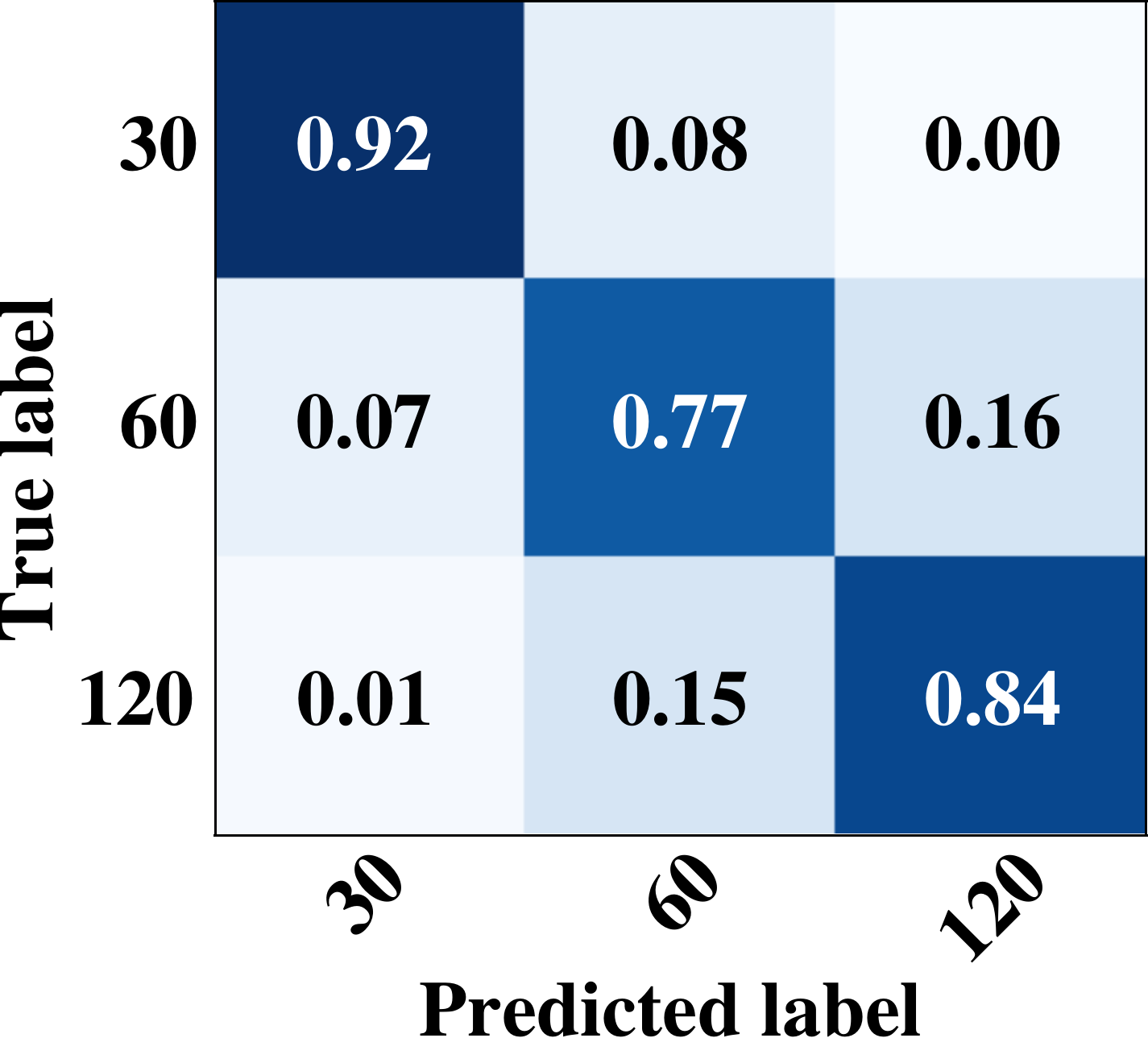}\label{fig:clf-res-cm}}
\caption{Individual classifier and overall scheme confusion matrices for a 10-fold cross-validation training with their respective datasets.}
\label{fig:cm}
\end{figure}

Figures~\ref{subfig:cm-rf1} and~\ref{subfig:cm-rf2} shows the resulting confusion matrices of both \gls{RF} models, individually. For the \textit{120fps-FD} classifier, error rates of $20\%$ and $17\%$ can be observed for the \textit{120fps} and \gls{FD} classes respectively, which represent a good performance considering the \gls{VFR} classification problem and its imperfect ground-truth. Indeed, the frontier between annotating a sequence with a \textit{120fps} label and a \textit{60fps} label can be difficult to maintain consistent during the processing of the 360 videos of the training set. In addition, as detailed in \Section{\ref{subsec:ground-truth-problem}}, several sequences with high motion discontinuities have been separated into shots with different labels. Since these motion change frontiers could only be determined subjectively and not at a precise frame level, some dataset samples, i.e, 4-frame chunks, located at these frontiers could have been annotated with incorrect labels. Therefore, the error rate is likely to be over-estimated, leading to a visible motion artifacts rate lower than the observed $20\%$. For the \textit{60fps-30fps} model, correct prediction rates are respectively $83\%$ and $91\%$ for the \textit{30fps} and \textit{60fps} classes. Critical errors, defined as critical frame-rate under-estimation, represent $9\%$ of the \textit{60fps} class samples. This rate can be problematic since the under-estimation with a frame-rate of 30 fps could lead to severe visible motion artifacts. However, the same aforementioned remark concerning imperfect ground truth applies to the training dataset of the \textit{60fps-30fps} model. The proportion of frame-rate over-estimation errors, equal to $17\%$, does not impact the visual quality and is thus not as prejudicial as the critical errors.

The overall VFR prediction scheme confusion matrix, obtained by combining the cross-validation validation-fold predictions of both models, is depicted in \Figure{\ref{fig:clf-res-cm}}. It is important to note that only a subset of randomly chosen \textit{120fps} class samples has been used to compute the overall prediction scheme confusion matrix so that the three classes have the same number of samples. The observed performance is consistent with the individual \gls{RF} model prediction results, with good probabilities of correct prediction. The only significant change is the $69\%$ correct prediction rate for the \textit{60fps} class, which can be explained by the fact that it is the intermediate class, thus sharing characteristics with the other two classes which makes the discrimination of the class samples harder to generalize. In addition, the \textit{extreme} errors, i.e. the critical under-estimation of a \textit{120fps} sample with a \textit{30fps} predicted label or the exact inverse, are rarely occurring with rates of $3\%$ and $1\%$, respectively. This tends to bolster the hypothesis on the ground truth being imperfect due to possibly unstable/blurry annotation frontiers between adjacent labels. If this hypothesis is correct, the combined prediction model should lead to \gls{VFR} output video sequences visually identical to the \gls{HFR} input. However, the compression and encoding complexity gains should be slightly lower than with ground truth labels. The next section aims at verifying this statement.

\section{Results and Analysis}
\label{sec:results} 

Before analyzing the coding performance of the \gls{VFR} coding scheme, the visual quality of the output \gls{VFR} video must be evaluated to assess whether the \gls{RF} model frame-rate decisions are preserving the perceptual quality compared to the \gls{HFR} source video. This section first describes the characteristics of the test set sequences and the chosen subjective evaluation methodology. Then, the results of the subjective tests are detailed and discussed for both uncompressed and compressed \gls{VFR} videos. Finally, the coding performance of \gls{VFR} coding scheme is presented in terms of bit-rate savings and complexity reduction.

\subsection{Specific Test Datasets and Subjective Tests Motivations}
\label{subsec:results-datasets}

A total of 15 sequences has been selected to validate the performance of the \gls{VFR} model. These sequences are unknown to the model, i.e. they have not been used during the cross-validation training of both binary \gls{RF} classifiers. The input frame-rate is 120 fps for all test sequences and their duration ranges between 9 and 13 seconds. Source content with a 3840x2160 original resolution have been downsampled to the 1920x1080 resolution with Lanczos3 filers~\cite{duchon1979lanczos} to ensure consistency during the subjective test. 

\begin{figure}[!t]
\centering
\includegraphics[width=0.9\linewidth]{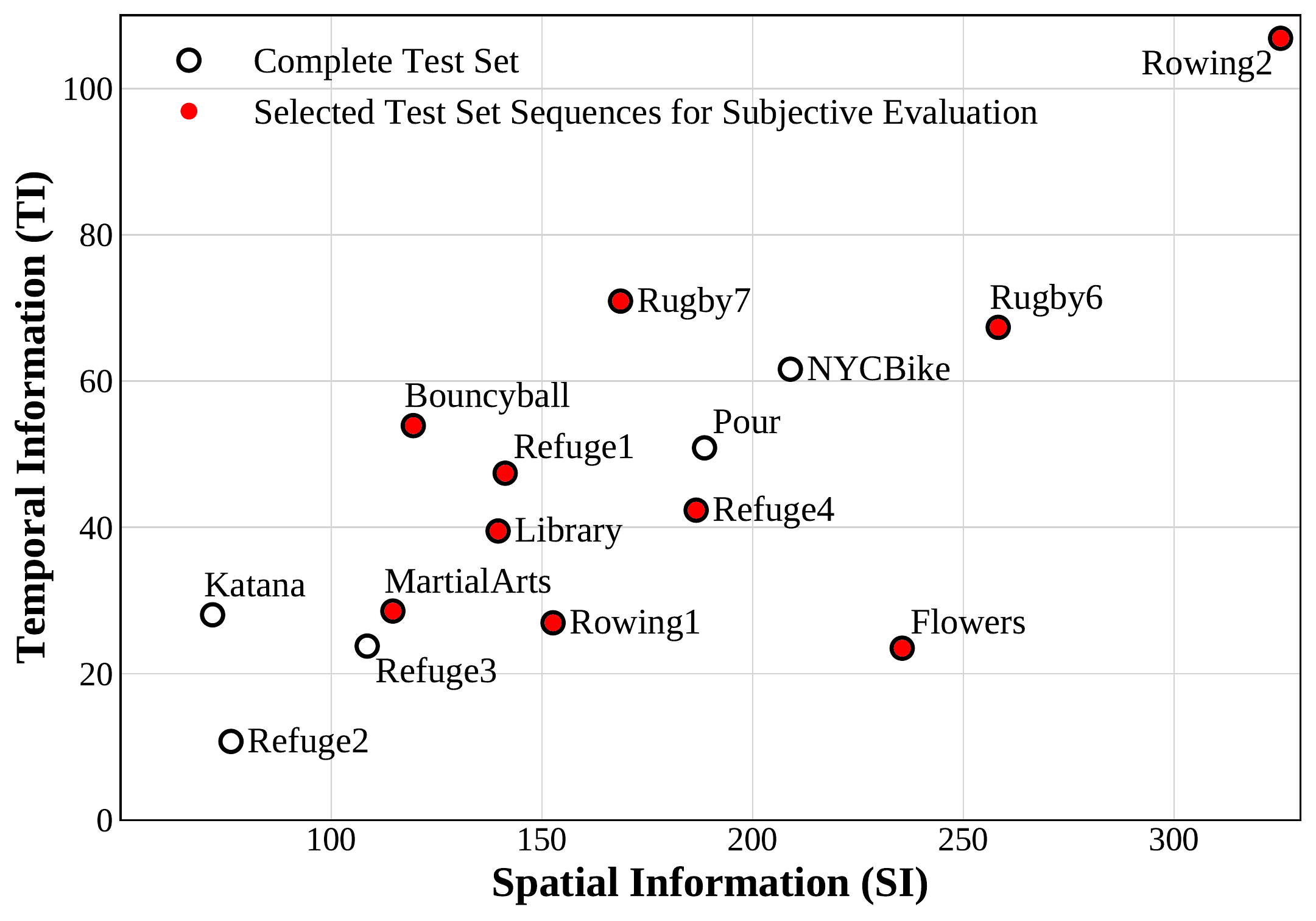}
\caption{SI-TI characteristics for test sequences of the three considered sets.}
\label{fig:si-ti-val}
\end{figure}

The test set sequences have been selected from various sources to cover a wide range of spatio-temporal characteristics both in terms of temporal and spatial information (SI and TI), as recommenced in ~\cite{recP910} and shown in \Figure{\ref{fig:si-ti-val}}. They also depict several use-cases, including sporting events and movie-type clips, in addition to the more common natural video content.

\begin{figure}[!t]
\centering
\includegraphics[width=0.42\linewidth]{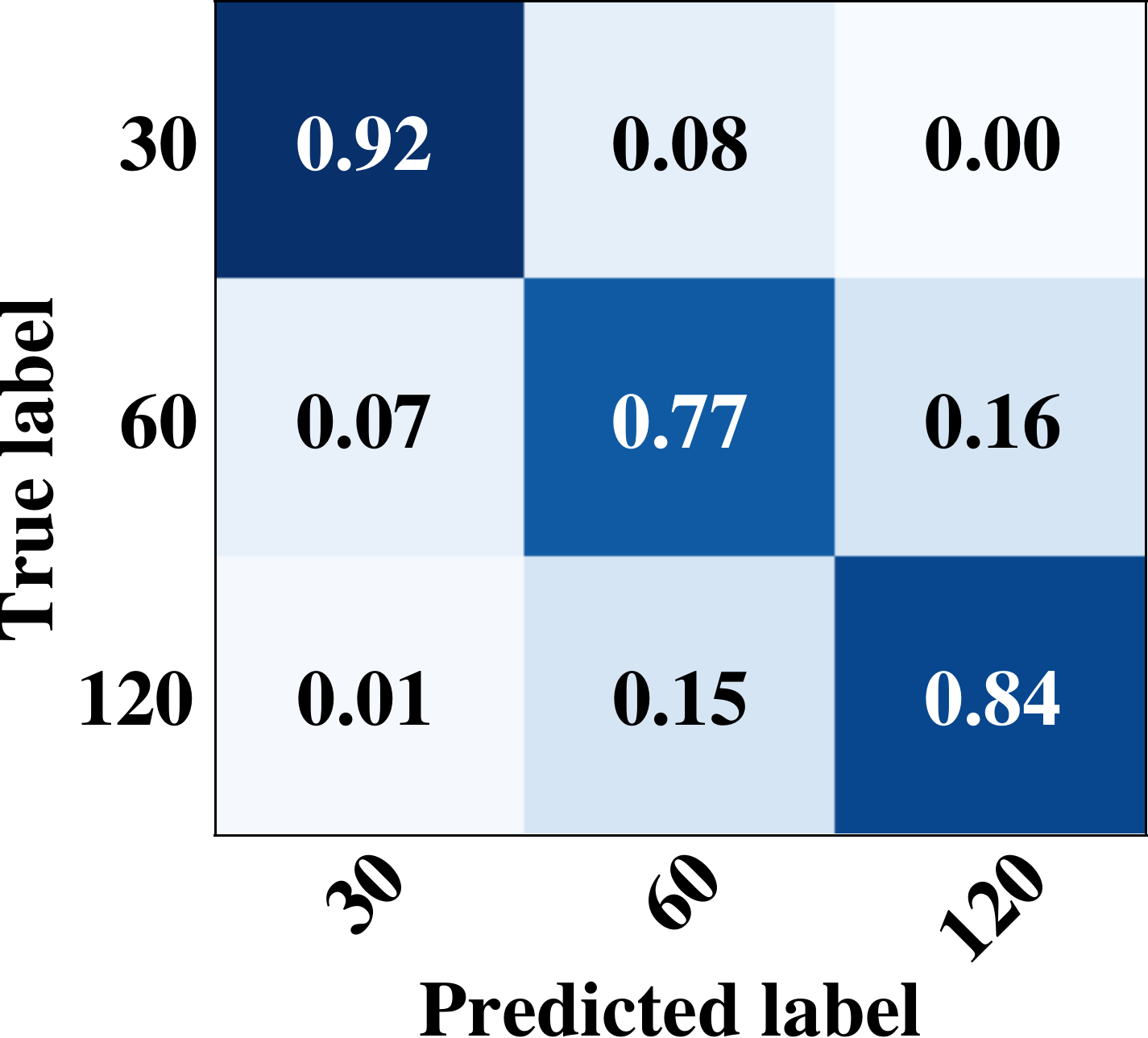}
\caption{Cascaded RF model prediction performance on test set.}
\label{fig:cm-val}
\end{figure}

In order to generate the final \gls{VFR} model, both \gls{RF} classifiers have been retrained with their respective whole datasets as well as their feature sets and hyper-parameters determined via cross-validation. The prediction results for the test set are depicted in \Figure{\ref{fig:cm-val}}. As can be observed, correct prediction rates reach $92\%$, $77\%$ and $84\%$ for the \textit{30fps}, \textit{60fps} and \textit{120fps} classes, respectively, showing the capacity of the model to generalize the \gls{VFR} classification problem to unknown data. The slightly better prediction results for the test set compared to the cross-validation predictions presented in Section~\ref{subsec:res-classif} may be explained by the more accurate ground truth labels obtained for the test set, thus minimizing the labeling issue previously raised. In addition, the low amount of samples falsely predicted with a \textit{30fps} class label should lead to a good perceptual quality of the \gls{VFR} output videos, very close to the \gls{HFR} source content. 

To verify this statement, a subjective test comparing the uncompressed \gls{HFR} and \gls{VFR} videos has been designed. The \textit{30fps} and \textit{60fps} versions, obtained by frame decimation, have also been introduced in the subjective evaluation to assess the interest of variable frame-rate compared to systematic temporal downsampling in terms of perceived quality. 

\subsection{Subjective Evaluation Methodology}
\label{subsec:results-methodo}

The considered subjective evaluation aims at assessing the effect of a system, here the \gls{VFR} coding scheme, on the visual quality. For this kind of test, the \gls{ITU}-R BT.500-13 recommendation~\cite{recBT500} proposes the \gls{DSCQS} method, which consists in showing the observer pairs of videos - the un-processed source content and the same sequence processed with the system under test - and asking the observer to rate the quality of both sequences. The grading scale is a continuous vertical scale divided into 5 equal parts corresponding to the common 5-level \gls{ITU}-R quality labels: \textit{Excellent}, \textit{Good}, \textit{Fair}, \textit{Poor} and \textit{Bad}.

\begin{figure}[!t]
\centering
\includegraphics[width=0.8\linewidth]{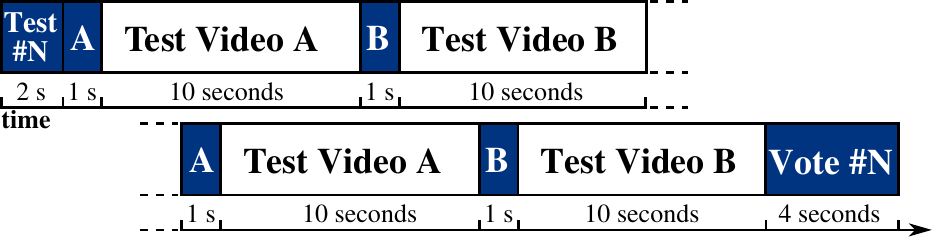}
\caption{Subjective test BTC structure for DSCQS evaluation method.}
\label{fig:btc-val}
\end{figure}

For each test session, a series of video pairs is presented to the observer in a random order, to distribute the degrees of quality impairments over the entire session. Each pair of videos is internally random, i.e. the observer is not aware of the position of the reference un-processed video (A or B), which is presented twice, successively. \Figure{\ref{fig:btc-val}} depicts the structure of a \gls{BTC} presenting a pair of videos to assess. As can be observed, each \gls{BTC} begins with a 2-second message indicating the id number of the current test point and ends with a message asking to vote. In addition, each display of a 10-second sequence is preceded by a 1-second message indicating if the following video is A or B on the answer sheet, making the total duration of a \gls{BTC} equal to 50 seconds. 

\begin{figure*}[!t]
\centering
\includegraphics[width=0.75\linewidth]{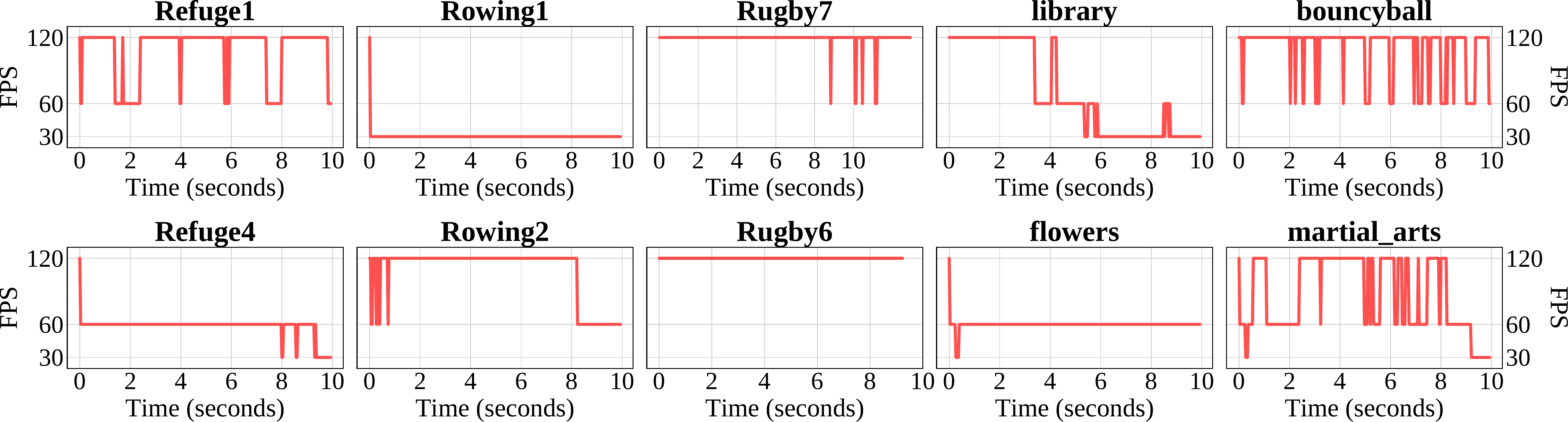}
\caption{Frame-rate decisions of the VFR algorithm for test set sequences.}
\label{fig:vfr-decisions}
\end{figure*}

A total number of 10 sequences have been selected within the test set for the subjective test, as indicated in \Figure{\ref{fig:si-ti-val}}. The sequence set has been formed to cover a wide range of spatio-temporal characteristics and content types. For each sequence, 4 frame-rate pairs have been evaluated by the observers: \textit{120fps vs 120fps}, \textit{120fps vs \gls{VFR}}, \textit{120fps vs 60fps} and \textit{120fps vs 30fps}. Therefore, a total of 40 \glspl{BTC} were presented to each observer, randomly divided into two 20-minute sessions separated by a 10-minute break. \gls{VFR} video sequences have been obtained using the predicted frame-rates resulting from the proposed \gls{VFR} model. \Figure{\ref{fig:vfr-decisions}} depicts the evolution of frame-rate decisions over the duration of each sequence of the subjective evaluation test set. As can be observed, the predicted frame-rates are highly dependent on the test sequence, as expected considering the wide range of spatial and temporal information characteristics for the selected sequence set. In addition, the predicted frame-rates also vary over-time for most of the test sequences, demonstrating the interest of the 4-frame level of granularity proposed for frame-rate decisions.

The test was conducted in a controlled laboratory environment, with a viewing distance fixed to 3 times the screen height. A 65-inch LG OLED B6 display with HFR capabilities and peak luminance of 340 cd/m$^2$ has been used for both subjective tests. During the whole duration of the tests, all internal post-processing were disabled to avoid any impact on the perceived quality. Each test sequence in raw format (YUV 4:2:0 and 8-bit precision) has been encoded using the \textit{libx265} encoder at 100 Mbps in order to be presented to the TV set via USB3 interface. Special care has been taken to ensure that the encoding needed for display did not introduce any `coding' artifacts. A total of 19 participants took part in the subjective test. They were aged between 20 and 53 with (corrected-to) normal vision acuity and color vision. A post-screening analysis of the results has been carried out, according to the method described in \gls{ITU}-R Rec. BT.500-13, to detect and reject the outliers before computing the \gls{MOS} values.

\subsection{Subjective Visual Quality Results}
\label{subsec:results-subj}

\begin{figure*}[!t]
\centering
\includegraphics[width=0.55\linewidth]{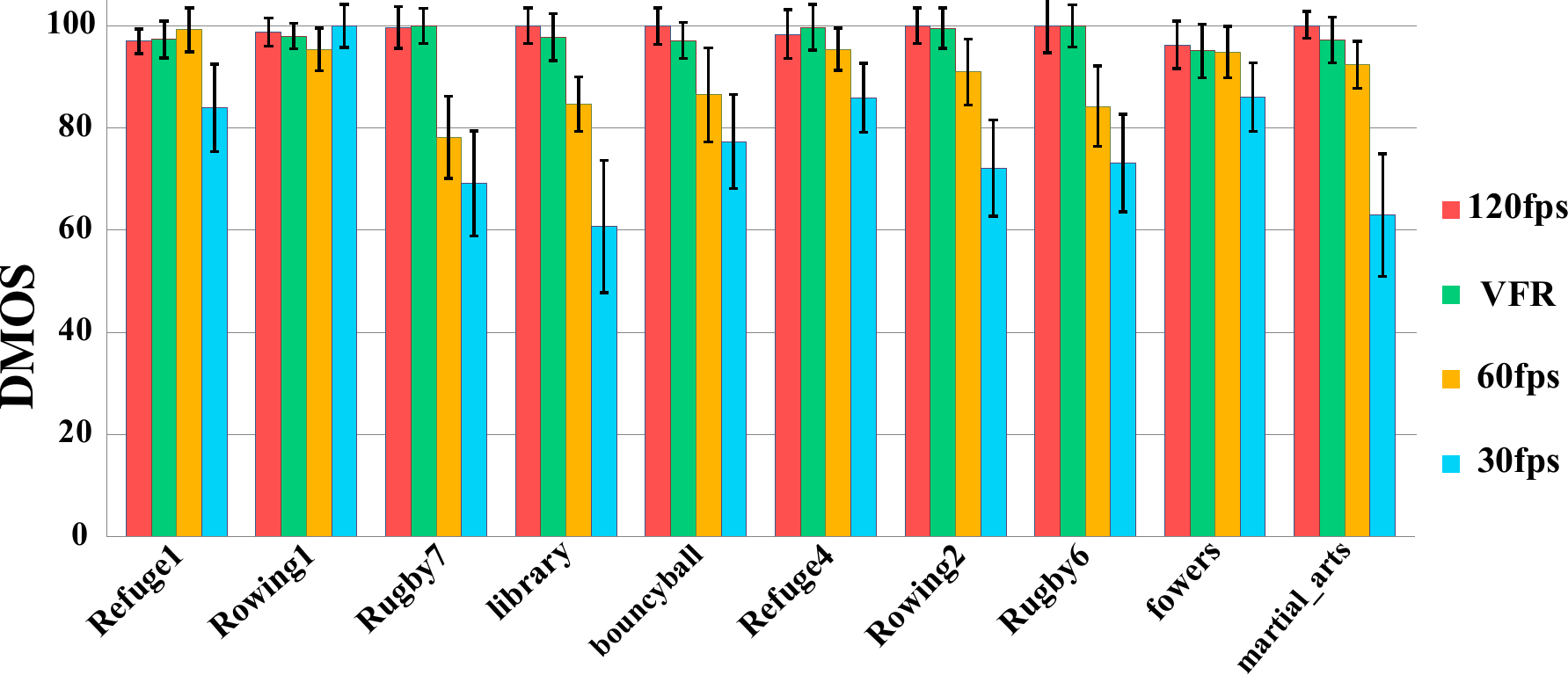}
\vspace*{-0.25em}\caption{Mean Opinion Score values with 95\% confidence intervals for test set sequences and subjectively tested frame-rates.}
\label{fig:vfr-results}

\end{figure*}

\Figure{\ref{fig:vfr-results}} shows the results of the subjective test carried out to demonstrate the interest of variable frame-rate and evaluate the perceived quality of the proposed \gls{VFR} model output. For each sequence of the test set previously presented, the \gls{DMOS} values, computed using Equation~(\ref{eq:mos-value}), of each tested frame-rate are depicted together with their associated $95\%$ \glspl{CI}. Since none of the participants were flagged as outliers after the post-screening analysis, the presented \gls{DMOS} values have been obtained using the results from the 19 participants.
\begin{equation}
DMOS_{f}(s) = 100 - \dfrac{1}{N} \, \sum^{N}_{n=1} S_{n,120fps}(s)-S_{n,f}(s),
\label{eq:mos-value}
\end{equation}
with $N$ the total number of valid participants, $N=19$ in this test, $DMOS_{f}(i)$ the \gls{DMOS} value for sequence $s$ at the tested frame-rate $f$, $f \in \{120fps,\,VFR,\,60fps,\,30fps\}$. The pair $(S_{n,120fps}(s),S_{n,f}(s))$ represents the scores attributed to sequence $s$ at respectively the hidden reference \textit{120fps} frame-rate and tested frame-rate $f$, i.e. both videos of a given \gls{BTC}, by the $n^{th}$ participant, $n \in \{1,..,N\}$.

The first statement that can be made by analyzing the results of the subjective test is that, as previously stated, the benefit brought by a frame-rate of 120 images per second compared to lower frame-rates is highly content-dependent. Indeed, for the sequences \textit{Rugby7}, \textit{library} and \textit{Rugby6}, there is a significant difference between the \gls{DMOS} values associated to the \textit{120fps} frame-rate and those of the \textit{60fps} and \textit{30fps} frame-rates. The same trend can be observed for the \textit{bouncyball}, \textit{Rowing2} and \textit{martials\_arts} sequences. However, for these sequences, the \glspl{CI} of the \textit{120fps} and \textit{60fps} \gls{DMOS} are overlapping, thus a significant difference between the perceived quality of the two frame-rates cannot be confidently guaranteed for these sequences. For other sequences, namely \textit{Refuge1}, \textit{Refuge4} and \textit{flowers}, the perceived quality of the \textit{120fps} and \textit{60fps} seem equivalent, with similar \gls{DMOS} and highly overlapping \glspl{CI}. Finally, the sequence \textit{Rowing1} shows no visual difference even with a frame decimation down to \textit{30fps}.

Comparing the perceived qualities of the \gls{VFR} model outputs with their source \gls{HFR} 120 fps counterpart, the \gls{DMOS} values of both configurations appear to be equivalent for every sequence. This trend highlights the interest of variable frame-rate with its capacity to adapt to the quantity of movement possibly varying over time. For instance, the \textit{library} sequence opens on a camera panning with a gradually slowing speed which then stops at the middle of the sequence on a stationary top spinning at high speed. The first part of the video requires 120 fps to correctly portray the camera panning, while lower frame-rates can be used without introducing artifacts as the speed of the camera gradually drops. For this sequence, participants attributed significantly lower scores to the \textit{60fps} and \textit{30fps} frame-rates due to the important motion artifacts present in the first part of the video at these frame-rates. On the contrary, the \gls{VFR} model correctly lowers the frame-rate when the content permits it, resulting in a score identical to the one attributed to the source \gls{HFR} video. However, despite highly correlated \gls{DMOS} values and overlapping \glspl{CI}, there is still a chance that the perceived qualities of the compared frame-rates are actually different.

\begin{figure*}[!t]
\centering
\subfloat[Refuge1]{\scriptsize
\begin{tabular}{c|ccc}
FPS & {120} & VFR & 60 \\ \hline%
\bigstrut VFR & {\bfseries\cellcolor{green!50}0.77}   & \cellcolor{gray!20}   & \cellcolor{gray!20}  \\ 
\bigstrut 60  & {\bfseries\cellcolor{green!50}0.26}   & {\bfseries\cellcolor{green!50}0.41}   & \cellcolor{gray!20}  \\ 
\bigstrut 30  & {\bfseries\cellcolor{red!40}0.02}   & {\bfseries\cellcolor{red!40}0.03}   & {\bfseries\cellcolor{red!40}0.00}  \\ \hline%
\end{tabular}
} \;%
\subfloat[Rowing1]{\scriptsize
\begin{tabular}{c|ccc}
FPS & {120} & VFR & 60 \\ \hline%
\bigstrut VFR & {\bfseries\cellcolor{green!50}0.96}   & \cellcolor{gray!20}   & \cellcolor{gray!20}  \\ 
\bigstrut 60  & {\bfseries\cellcolor{green!50}0.14}   & {\bfseries\cellcolor{green!50}0.14}   & \cellcolor{gray!20}  \\ 
\bigstrut 30  & {\bfseries\cellcolor{green!50}0.42}   & {\bfseries\cellcolor{green!50}0.44}   & {\bfseries\cellcolor{green!50}0.12}  \\ \hline%
\end{tabular}
} \;%
\subfloat[Rugby7]{\scriptsize
\begin{tabular}{c|ccc}
FPS & {120} & VFR & 60 \\ \hline%
\bigstrut VFR & {\bfseries\cellcolor{green!50}0.56}   & \cellcolor{gray!20}   & \cellcolor{gray!20}  \\ 
\bigstrut 60  & {\bfseries\cellcolor{red!40}0.00}   & {\bfseries\cellcolor{red!40}0.00}   & \cellcolor{gray!20}  \\ 
\bigstrut 30  & {\bfseries\cellcolor{red!40}0.00}   & {\bfseries\cellcolor{red!40}0.00}   & {\bfseries\cellcolor{red!40}0.01}  \\ \hline%
\end{tabular}
} \;%
\subfloat[library]{\scriptsize
\begin{tabular}{c|ccc}
FPS & {120} & VFR & 60 \\ \hline%
\bigstrut VFR & {\bfseries\cellcolor{green!50}0.56}   & \cellcolor{gray!20}   & \cellcolor{gray!20}  \\ 
\bigstrut 60  & {\bfseries\cellcolor{red!40}0.00}   & {\bfseries\cellcolor{red!40}0.00}   & \cellcolor{gray!20}  \\ 
\bigstrut 30  & {\bfseries\cellcolor{red!40}0.00}   & {\bfseries\cellcolor{red!40}0.00}   & {\bfseries\cellcolor{red!40}0.00}  \\ \hline%
\end{tabular}
}%\\
\subfloat[bouncyball]{\scriptsize
\begin{tabular}{c|ccc}
FPS & {120} & VFR & 60 \\ \hline%
\bigstrut VFR & {\bfseries\cellcolor{green!50}0.41}         & \cellcolor{gray!20}   & \cellcolor{gray!20}  \\ 
\bigstrut 60  & {\bfseries\cellcolor{red!40}0.01}   & {\bfseries\cellcolor{red!40}0.00}   & \cellcolor{gray!20}  \\ 
\bigstrut 30  & {\bfseries\cellcolor{red!40}0.00}   & {\bfseries\cellcolor{red!40}0.00}   & {\bfseries\cellcolor{red!40}0.01}  \\ \hline%
\end{tabular}
}\\[-0.5em]%
\subfloat[Refuge4]{\scriptsize
\begin{tabular}{c|ccc}
FPS & {120} & VFR & 60 \\ \hline%
\bigstrut VFR & {\bfseries\cellcolor{green!50}0.85}   & \cellcolor{gray!20}   & \cellcolor{gray!20}  \\ 
\bigstrut 60  & {\bfseries\cellcolor{green!50}0.27}   & {\bfseries\cellcolor{green!50}0.27}   & \cellcolor{gray!20}  \\ 
\bigstrut 30  & {\bfseries\cellcolor{red!40}0.02}   & {\bfseries\cellcolor{red!40}0.01}   & {\bfseries\cellcolor{red!40}0.03}  \\ \hline%
\end{tabular}
} \;%
\subfloat[Rowing2]{\scriptsize
\begin{tabular}{c|ccc}
FPS & {120} & VFR & 60 \\ \hline%
\bigstrut VFR & {\bfseries\cellcolor{green!50}0.45}   & \cellcolor{gray!20}   & \cellcolor{gray!20}  \\ 
\bigstrut 60  & {\bfseries\cellcolor{red!40}0.02}   & {\bfseries\cellcolor{red!40}0.01}   & \cellcolor{gray!20}  \\ 
\bigstrut 30  & {\bfseries\cellcolor{red!40}0.00}   & {\bfseries\cellcolor{red!40}0.00}   & {\bfseries\cellcolor{red!40}0.00}  \\ \hline%
\end{tabular}
}\;%
\subfloat[Rugby6]{\scriptsize
\begin{tabular}{c|ccc}
FPS & {120} & VFR & 60 \\ \hline%
\bigstrut VFR & {\bfseries\cellcolor{green!50}0.26}         & \cellcolor{gray!20}   & \cellcolor{gray!20}  \\ 
\bigstrut 60  & {\bfseries\cellcolor{red!40}0.00}   & {\bfseries\cellcolor{red!40}0.00}   & \cellcolor{gray!20}  \\ 
\bigstrut 30  & {\bfseries\cellcolor{red!40}0.00}   & {\bfseries\cellcolor{red!40}0.00}   & {\bfseries\cellcolor{red!40}0.00}  \\ \hline%\bottomrule
\end{tabular}
}\;%
\subfloat[flowers]{\scriptsize
\begin{tabular}{c|ccc}
FPS & {120} & VFR & 60 \\ \hline
\bigstrut VFR & {\bfseries\cellcolor{green!50}0.99}   & \cellcolor{gray!20}   & \cellcolor{gray!20}  \\ 
\bigstrut 60  & {\bfseries\cellcolor{green!50}0.56}   & {\bfseries\cellcolor{green!50}0.58}   & \cellcolor{gray!20}  \\ 
\bigstrut 30  & {\bfseries\cellcolor{red!40}0.00}   & {\bfseries\cellcolor{red!40}0.00}   & {\bfseries\cellcolor{red!40}0.01}  \\ \hline
\end{tabular}
}   
\subfloat[martial\_arts]{\scriptsize
\begin{tabular}{c|ccc}
FPS & {120} & VFR & 60 \\ \hline
\bigstrut VFR & {\bfseries\cellcolor{green!50}0.13}   & \cellcolor{gray!20}   & \cellcolor{gray!20}  \\ 
\bigstrut 60  & {\bfseries\cellcolor{red!40}0.00}   & {\bfseries\cellcolor{red!40}0.03}   & \cellcolor{gray!20}  \\ 
\bigstrut 30  & {\bfseries\cellcolor{red!40}0.00}   & {\bfseries\cellcolor{red!40}0.00}   & {\bfseries\cellcolor{red!40}0.00}  \\ \hline
\end{tabular}
}
\caption{\textit{p-value} probabilities resulting from two-sample unequal variance bilateral Student's t-test on \gls{MOS} values for each pair of tested frame-rates and each test set sequence. \textit{$p\geq0.05$ (green) means there is no significant difference between the \gls{MOS} value of the row and column frame-rate labels while $p<0.05$ (red) indicates that the \gls{MOS} value of the row frame-rate label is significantly lower than the \gls{MOS} value of the column frame-rate label.}}
\label{fig:student-test}
\end{figure*}

To confirm these observations and confidently state that the \gls{VFR} model output perceived quality is the same as for the source \gls{HFR} content, a more rigorous analysis can be performed using a two-sample unequal variance Student's t-test with a two-tailed distribution (also called Welsch's t-test). This test allows to determine if indeed the perceived qualities given by the \gls{MOS} values of each pair of tested frame-rates are ``significantly'' different or not. In this case, the null hypothesis, $H_0$, would be that the tested frame-rate $f_{test}$ has the same perceived quality as the considered reference frame-rate $f_{ref}$. The alternate hypothesis, $H_a$, would be that there is a difference between the perceived qualities of $f_{test}$ and $f_{ref}$. In order to test the similarity for each possible pair of frame-rates, the possible values for both frame-rates are: $f_{test} \in \{VFR,\,60fps,\,30fps\}$ and $f_{ref} \in \{120fps,\,VFR,\,60fps\}$. 

First, considering the sample populations from the scores attributed to a sequence $s$ at the two frame-rates $f_{test}$ and $f_{ref}$ being compared, the t-statistic $t_{f_{test},f_{ref}}(s)$ can be used, expressed as follows
\begin{equation}
t_{f_{test},f_{ref}}(s) = \dfrac{\overline{S}_{f_{test}}(s)-\overline{S}_{f_{ref}}(s)}{\sqrt{\dfrac{\sigma^2_{f_{test}}(s)}{N_{f_{test}}}+\dfrac{\sigma^2_{f_{ref}}(s)}{N_{f_{ref}}}}},
\label{eq:t-stat}
\end{equation}
with $\overline{S}_{f_i}(s)$, $\sigma^2_{f_i}(s)$, $N_{f_i}$ the sample mean, sample variance and sample population size for frame-rate $f_i$, $i \in \{test,\,ref\}$. In this test, $N_{f_{test}} = N_{f_{ref}} = N$, the number of observers that took part in the subjective test.

Then, by approximating the t-statistic with a Student's t-distribution, a value $p$, which indicates the degree of correlation between the means of the two sample populations, can be computed from the t-statistic. The higher the \textit{p-value} is, the more significant the similarity between the distributions of the two populations is. A \textit{p-value} lower than $0.05$ indicates that there is statistical significance that the tested frame-rate $f_{test}$ has a different perceived quality compared to the considered reference frame-rate $f_{ref}$. Indeed, in this case, there is a low probability of committing a type-I error, i.e. rejecting the null hypothesis when it is true, meaning that the null hypothesis can be confidently rejected. On the contrary, if the \textit{p-value} is greater than or equal to $0.05$, the null hypothesis cannot be safely rejected and both frame-rates can be considered to have the same perceived quality. 

Finally, the \textit{p-value} does not give information on the probability of committing a type-II error, i.e. a failure to reject the null hypothesis when the alternate hypothesis is true, which is thus still a possibility. To ensure a low type-II error probability, and thus a statistically powerful test, the power $\beta$ of the statistical test must be lower than $0.2$. The power $\beta$ has been computed for each possible pair of tested and reference frame-rates, resulting in an average $\beta$ value of $0.044$, showing that there is a lower than $5\%$ chance, on average, to commit a type-II error. Therefore, the similarity assessment for each pair of possible frame-rates can be only based on the \textit{p-values} resulting from the Student's t-test.

\Figure{\ref{fig:student-test}} depicts the \textit{p-values} computed for each sequence and each possible frame-rate combination. Green-colored cells show the frame-rate pairs for which the associated \textit{p-value} is greater than $0.05$. Since every \textit{\gls{VFR}} vs \textit{120fps} comparison falls within this category, it can be confidently concluded that the perceived quality of the \gls{VFR} model output video is always the same as the original 120 fps frame-rate. This confirms that the under-estimated frame-rate predictions, identified in the confusion matrix depicted in \Figure{\ref{fig:cm-val}}, do not impact the perceived quality of the \gls{VFR} videos. This also tends to validate the hypothesis made while analyzing training errors, stating that the ground truth is imperfect due to the coarse-grained nature of the ground truth annotations. Indeed, with its fine-grained decisions, the \gls{VFR} model is capable of capturing smaller variations of critical frame-rates, thus resulting in predictions different from the ground truth, which are identified as prediction errors.

\subsection{Compression Efficiency and Complexity Reduction}
\label{subsec:results-obj}

In order to evaluate the impact of \gls{VFR} on coding performance, both the source \gls{HFR} videos and \gls{VFR} model outputs have been encoded using the \gls{HEVC} reference software encoder HM16.12~\cite{hm16-12}. The encoder was configured to use the \gls{HEVC} \gls{CTC} in \gls{RA} configuration with a \gls{GOP} size of 16 pictures and an intra-period of approximately 1 second to match with the considered broadcasting use-case. The quantization parameter was set to $QP=\{22,27,32,37\}$ to cover a wide range of bit-rates and applications.

\begin{figure}[!t]
\centering
\includegraphics[width=\linewidth]{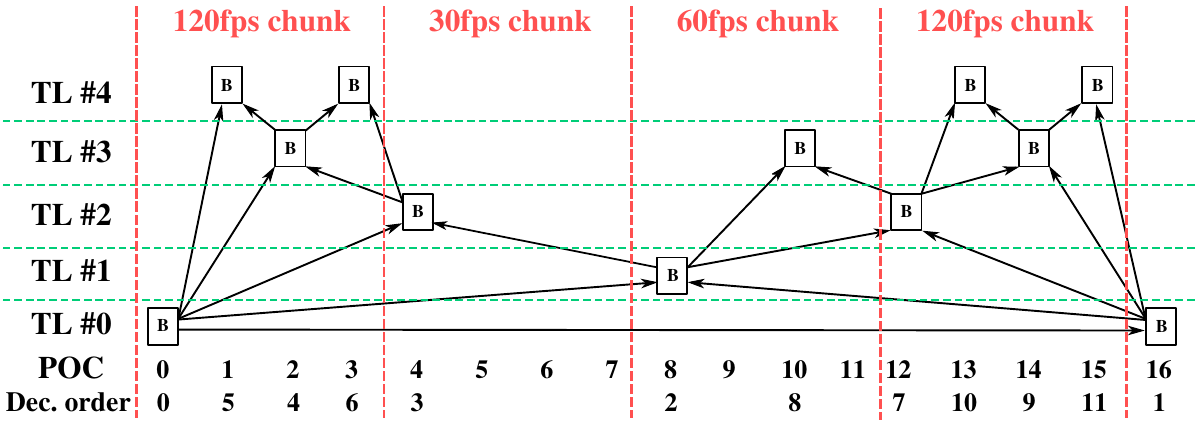}
\caption{Example of GOP structures of size 16 for a) source HFR 120 fps content and b) VFR with different frame-rates for each 4-frame chunk.}
\label{fig:gop}
\end{figure}

For the \gls{VFR} encodings, the \gls{HEVC} reference software encoder has been modified to handle the critical frame-rate decision coming from the proposed \gls{VFR} module for each chunk of 4 input frames. \Figure{\ref{fig:gop}} depicts the \gls{GOP} structures needed to be supported in order to encode a \gls{VFR} video sequence. Thanks to the built-in support of temporal scalability, removing the frames from upper \glspl{TL} does not break the coding dependencies. The \gls{VFR} \gls{GOP} structure is thus enabled in the reference software by simply skipping frames within the core encoding loop depending on the given frame-rate decision fed to the encoder and the current \gls{POC}. This results in a \gls{VFR} encoding with a lower bit-rate and reduced coding complexity while producing a bitstream decodable by the reference software decoder without any modification.

The bit-rate savings presented in this performance evaluation assume that, with the same \gls{QP}, the perceived quality of the \gls{VFR} decoded video is the same as the decoded original \textit{120fps} video, as was demonstrated for the \gls{VFR} and \textit{120fps} uncompressed inputs. Indeed, on one hand, any frame decoded from the \gls{VFR} bitstream is exactly the same as its corresponding frame in the \textit{120fps} decoded video due to the use of identical \gls{GOP} structures.  On the other hand, the validity of the \gls{VFR} frame-rate decisions on compressed data has been verified through an expert subjective test. This test aimed at evaluating the visual quality of the compressed \textit{120fps} and \gls{VFR} sequences, independently, for a subset of the \gls{VFR} test set. The protocol used is the standard \gls{DCR} method~\cite{recP910} with an 11-grade scale. This subjective test resulted in the same \gls{MOS} values for both the \textit{120fps} and \gls{VFR} decoded sequences, at every bit-rate tested, showing that the critical frame-rate decisions obtained with the \gls{VFR} model trained on uncompressed data remain valid for compressed content. The subjective test also showed that the frame-rate could be further reduced in some cases due to the removal of details, by the encoding process, that justified a higher frame-rate in the original uncompressed video. The coding performance of the proposed model could thus be slightly improved by training it on compressed data. However, such an improvement would require the annotation of the entire database at several \glspl{QP}, which would be a very time consuming task for a small coding gain. The encoding results presented in this work are thus obtained using the model trained on uncompressed videos.

Table~\ref{tab:perf-enc} summarizes the performance results for the \gls{VFR} encodings compared to regular \textit{120fps} HEVC encodings, in terms of both bit-rate savings and encoding complexity reduction for the 15 sequences of the objective evaluation test set. The proportion of frames dropped by the \gls{VFR} coding scheme are also added for information. Results for both the \gls{VFR} model and ground truth decisions are presented to compare the performance at two different levels of granularity. 

With the \gls{VFR} model decisions, the \gls{VFR} coding scheme offers $4.3\%$ bit-rate savings on average, ranging from $0\%$ to $15.4\%$ for sequences where 120 and 30 frames per second are chosen for the whole sequence, respectively. For sequences with mostly \textit{60fps} chosen or with temporally varying decisions, the bit-rate savings are generally around $4\%$. These bit-rate savings are not equal to the proportion of frames dropped by the \gls{VFR} model due to the significantly lower amount of bits  used to encode the frames of the upper \glspl{TL}. Indeed, upper \gls{TL} frames are coded using higher quantization steps and greatly benefit from the inter-picture predictions of the \gls{RA} coding configuration. Thus, the amount of transmitted quantized residuals is lower for these frames, especially if the motion is easily predictable and if high spatial details are not present in the source content. For the complexity reduction brought by the \gls{VFR} coding scheme, the results are close to the amount of frames dropped, with an average encoding complexity reduction of $28\%$, ranging from $0\%$ to $70\%$. The per-sequence results follows the same trend as for bit-rate savings but with higher gain variations. The difference between the complexity reduction and frames dropped results mainly comes from the slightly reduced coding complexity of the upper \gls{TL} frames compared to the kept frames of the lower \glspl{TL}. Indeed, a higher number of residual coefficient to binarize and process with the entropy coding engine increases the encoding time. For the decoding complexity, the detailed results are not presented in this paper but the observed gains are highly similar to those observed for the encoding side, with an average decoding complexity reduction of $27.4\%$ for the \gls{VFR} coding scheme.

\begin{table}[t]
\centering
\caption{VFR HEVC encoding performance compared to 120fps HEVC encodings for VFR predicted labels (Model) and ground truth (G-T) labels on the test set.}
\label{tab:perf-enc}
\scriptsize
\begin{tabular}{l|cc|cc|cc}
\multirow{3}{*}{Sequence} & \multicolumn{2}{c|}{\multirow{2}{*}{bit-rate savings}}  & \multicolumn{2}{c|}{Enc. Time} &  \multicolumn{2}{c}{Frames} \\ 
                          &  & & \multicolumn{2}{c|}{Reduction} &  \multicolumn{2}{c}{Dropped} \\ 
                          & \scriptsize{Model} & \scriptsize{G-T}  & \scriptsize{Model} & \scriptsize{G-T} & \scriptsize{Model} & \scriptsize{G-T}  \\ 
\toprule
Refuge1                   &    -1.1 \%       &      -4.9 \%     &        7.8 \%     &    39 \%    &   10 \%   & 50 \%   \\ 
Rowing1                   &    -9.3 \%       &      -9.3 \%     &        60 \%      &    60 \%    &   75 \%   & 75 \%   \\ 
Rugby7                    &    -0.1 \%       &       0.0 \%     &         0.6 \%    &     0.0 \%  &   0.9 \%  & 0.0 \%  \\ 
library                   &    -5.0 \%       &      -4.8 \%     &        39 \%      &    37 \%    &   42 \%   & 41 \%   \\ 
bouncyball                &    -2.5 \%       &       0.0 \%     &         6.7 \%    &     0.0 \%  &   8.7 \%  & 0.0 \%  \\ 
Refuge4                   &    -3.2 \%       &      -3.5 \%     &        47 \%      &    49 \%    &   52 \%   & 55 \%   \\ 
Rowing2                   &    -0.6 \%       &      -0.4 \%     &         6.1 \%    &     5.4 \%  &   9.7 \%  & 8.7 \%  \\ 
Rugby6                    &     0.0 \%       &       0.0 \%     &         0.0 \%    &     0.0 \%  &   0.0 \%  & 0.0 \%  \\ 
flowers                   &    -4.1 \%       &      -4.1 \%     &        41 \%      &    40 \%    &   50 \%   & 50 \%   \\ 
martial\_arts             &    -4.0 \%       &      -0.5 \%     &        23 \%      &     5.9 \%  &   28 \%   & 7.6 \%  \\ 
Katana                    &    -5.9 \%       &      -1.2 \%     &        35 \%      &    11 \%    &   38 \%   & 13 \%   \\ 
NYCBike                   &    -6.2 \%       &      -5.6 \%     &        27 \%      &    25 \%    &   32 \%   & 29 \%   \\ 
pour                      &    -1.6 \%       &      -1.6 \%     &        11 \%      &    11 \%    &   16 \%   & 15 \%   \\ 
Refuge2                   &   -15 \%         &      -14.6 \%     &        70 \%     &    68 \%    &   74 \%   & 71 \%   \\ 
Refuge3                   &    -5.8 \%       &      -5.6 \%     &        53 \%      &    52 \%    &   58 \%   & 56 \%   \\ 
\midrule
\textbf{Average}          &    \textbf{-4.3 \%}       &      \textbf{-3.7 \% }    &        \textbf{28 \%}      &    \textbf{27 \%}    &  \textbf{ 33 \%}   & \textbf{32 \%}  \\
\bottomrule
\end{tabular}
\end{table}

With the ground truth annotated frame-rates, the bit-rate savings and complexity reduction results are very close to the performance with the predicted frame-rates. This can be explained by the high correct prediction rate of the \gls{VFR} model on the test set. The results are only significantly different for some sequences. \textit{Refuge1} shows lower gains for the \gls{VFR} model output due to the over-estimation of the required frame-rate, i.e. alternation between \textit{120fps} and \textit{60fps} prediction while the annotated ground truth frame-rate is \textit{60fps} for the major part of the sequence. The opposite situation can be observed for the sequences \textit{bouncyball}, \textit{martial\_arts} and \textit{Katana}, where the \gls{VFR} model allows for lower frame-rates more frequently than the ground truth, thus resulting in higher gains.

\section{Conclusion}
\label{sec:conclusion}

In this paper, a new variable frame-rate coding scheme is proposed for broadcast delivery of \gls{HFR} (120 fps) contents. The proposed scheme incorporates a machine learning based \gls{VFR} model capable of dynamically adapting the frame-rate of the video before encoding and transmitting it to the end receiver. 

The \gls{VFR} model relies on several spatio-temporal features extracted from each frame of the input video to predict the optimal lowest artifact-free frame-rate through two cascaded binary \gls{RF} trained classifiers. The considered frame-rate adaptation is performed dynamically by choosing for each chunk of 4 consecutive input frames its associated critical frame-rate, among the three possible values: \textit{30fps}, \textit{60fps} or \textit{120fps}. The model achieves an average critical frame-rate correct prediction rate of $84\%$, while keeping the frame-rate under-estimations error rate below $8\%$. The visual quality of the generated \gls{VFR} videos has been carefully evaluated through formal subjective tests showing an identical perceived quality compared to the source \gls{HFR} content.

From a coding performance perspective, the proposed \gls{VFR} coding scheme provides average bit-rate savings of $4.3\%$ in addition to average complexity reductions of $28\%$ and $27.4\%$ at the encoding and decoding sides, respectively. {It should be noted that this work can be applied to other broadcast frame rates such us 25 and 50 and 100 \gls{fps} by adopting the proposed algorithm}.

The work proposed in this paper has been shown, at both the \gls{IBC} 2019 and \gls{NAB} Show 2019, through a real-time demonstration with both the input legacy \gls{HFR} and processed \gls{VFR} videos displayed synchronously on two \gls{HFR} screens to demonstrate the equivalence in perceived quality. The demonstration includes a real-time software implementation of the \gls{VFR} prediction model - 7.2 ms (138 fps) average runtime for feature computation (6.9 ms) and frame-rate prediction (0.3 ms) of HD sequences processed on a common consumer CPU.

The proposed solution is a practical candidate to lower the requirements for the broadcast delivery of the upcoming \gls{HFR} services of the \gls{DVB} \gls{UHD} second deployment phase. Additionally, thanks to the its hardware-friendly solution (feature computation based on well-known h.264 encoding and existing \gls{RF} hardware implementations\cite{van2012accelerating}), the proposed \gls{VFR} method can be considered by hardware video encoder manufacturers to enhance the quality of experience and reduce the energy fingerprint of their devices.


\section*{Acknowledgment}

The authors would like to thank Franck Chi, Maxime Peralta and Cl\'ement Brossard who also contributed to this project. 

\ifCLASSOPTIONcaptionsoff
  \newpage
\fi

% \IEEEtriggeratref{18}


\bibliographystyle{IEEEtran}
%\bibliography{IEEEabrv,refs}


\begin{IEEEbiography}[{\includegraphics[width=1in,height=1.25in,clip,keepaspectratio]{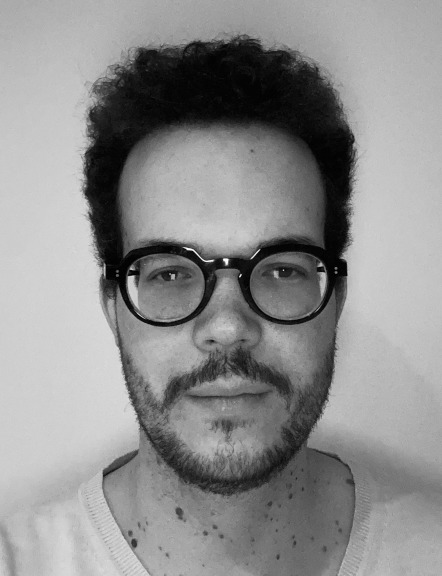}}]{Glenn Herrou}
received the  Dipl.-Ing. (M.Sc.) degree  in  electrical  and  computer  engineering and PhD in signal processing from the  Institut  National des  Sciences  Appliquées (INSA) de Rennes, France, in 2016 and 2019. From 2016 to 2019, he worked at the Institute of Research and Technology b\textless\textgreater com, Cesson-Sévigné, France, on projects focusing on adaptive spatio-temporal resolution for efficient video coding. Since 2020, he is a post-doctoral researcher in the VAADER team of the Institut d'Electronique et des Technologies du NuméRique (IETR), Rennes, France. His current research interests focus on video coding and applied Machine/Deep Learning.
\end{IEEEbiography}

\begin{IEEEbiography}[{\includegraphics[width=1in,height=1.25in,clip,keepaspectratio]{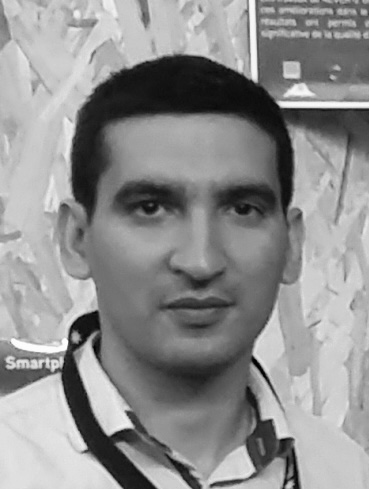}}]{Wassim Hamidouche}
received Master’s and Ph.D. degrees both in Image Processing from the University of Poitiers (France) in 2007 and 2010, respectively. From 2011 to 2013, he was a junior scientist in the video coding team of Canon Research Center in Rennes (France). He was a post-doctoral researcher from Apr. 2013 to Aug. 2015 with VAADER team of IETR where he worked under collaborative project on HEVC video standardisation. Since Sept. 2015 he is  an Associate Professor at INSA Rennes and a member of the VAADER team of IETR Lab. He has joined the Advanced Media Content Lab of b<>com IRT Research Institute as an academic member in Sept. 2017. His research interests focus on video coding and security of multimedia contents. He is the author/coauthor of more than one hundred and twenty (+120) papers at top journals and conferences in Image Processing, two MPEG standards, two patents, several MPEG contributions, public datasets and open source software projects.
\end{IEEEbiography}

\begin{IEEEbiography}[{\includegraphics[width=1in,height=1.25in,clip,keepaspectratio]{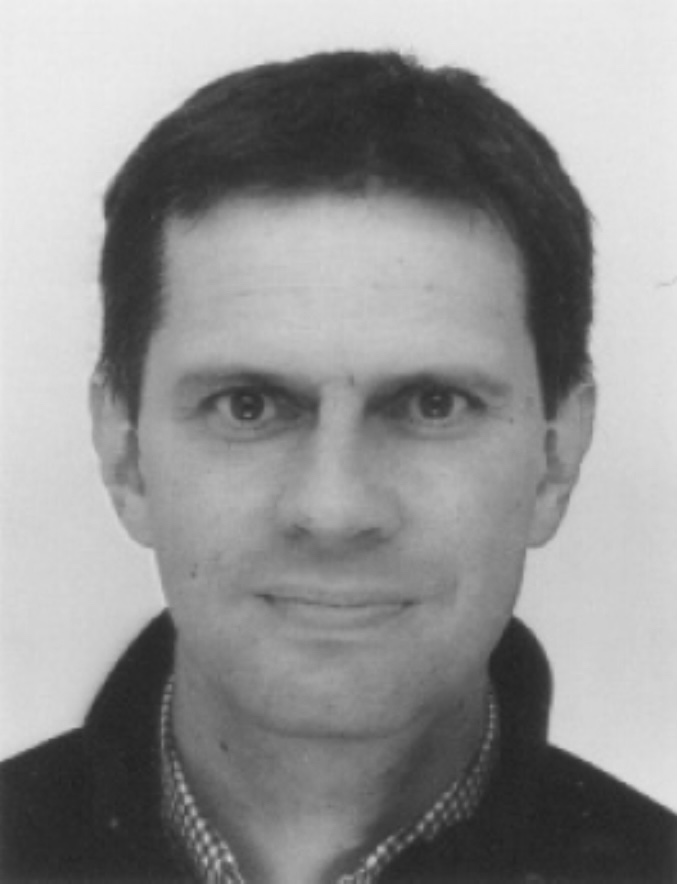}}]{J\'er\^ome Fournier} received the Ph.D. in signal and image processing from the University of Rennes, France, in 1995. He started his career at Philips in the field of video communications. In 1997, he joined Orange Labs (formerly France Telecom) and worked on video codecs like MPEG-4 Part 2 and H.264. From 2004 to 2012, Jérôme focused on the deployment of the Orange TV services, HDTV and stereoscopic 3DTV, as well as on innovative 3DTV depth-based video formats. From 2012 to 2018, he was mainly involved in the subjective evaluation and the ITU-R standardization of Ultra HD video formats including HDR and HFR features. Now, he is contributing to b\textless\textgreater com studies on topics like VFR, view synthesis and 8K.
\end{IEEEbiography}

\vfill
\pagebreak

\begin{IEEEbiography}[{\includegraphics[width=1in,height=1.25in,clip,keepaspectratio]{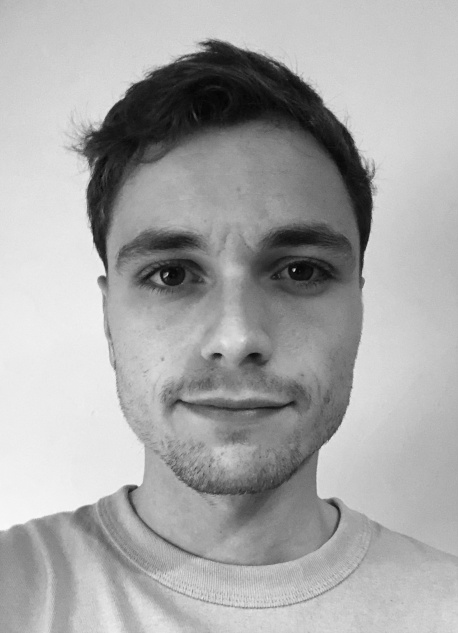}}]{Charles Bonnineau} received the Dipl.-Ing. degree in Computer Science at the Ecole Supérieure D'Ingénieurs de Rennes (ESIR) from the Université de Rennes 1, France, in 2018. He is currently a PhD Student in Signal and Image Processing jointly with the Institut d'Electronique et des Technologies du NuméRique (IETR), the Intitute of Research and Technology b\textless\textgreater com, and TDF. His current research include video processing and coding using deep-learning-based methods.   
\end{IEEEbiography}

\begin{IEEEbiography}[{\includegraphics[width=1in,height=1.25in,clip,keepaspectratio]{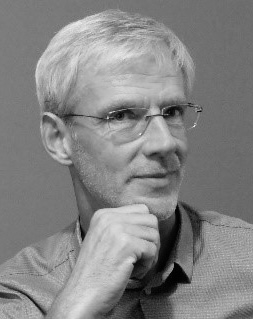}}]{Patrick Dumenil} received his Master's degree in Signal Processing from CentraleSupelec, Gif-sur-Yvette (France), in 1987. He started his career as a research engineer at Thomson Laboratoires Electronique, Rennes (France), where he developed new technologies for digital television. Then, he managed projects focused on designing first-generation digital television products and contributing to MPEG-2 and AVC standards. Since 2016, he leads the Codec Innovation team at Harmonic, Cesson-Sévigné (France), and is also involved in projects at the b\textless\textgreater com research institute. His current research focuses on developing Machine Learning solutions for codec efficiency improvement.
\end{IEEEbiography}

\begin{IEEEbiography}[{\includegraphics[width=1in,height=1.25in,clip,keepaspectratio]{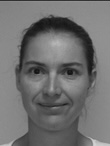}}]{Luce Morin} is a full professor at National Institute of Applied Sciences (INSA), Rennes, France, where she teaches computer science, image processing, and computer vision. She received the M.S. degree from ENSPS school in Strasbourg in 1989 and spent a 6 month internship at the NASA GSFC in Washington D.C. She then prepared a Ph.D. thesis supervised by professor Roger Mohr in the LIFIA laboratory, INP-Grenoble, on projective invariants applied to computer vision. From 1993 to 2008, she was an associate-professor at University of Rennes and a member of the Temics team in the IRISA/INRIA-Rennes laboratory. She is now a member of the Institut d'Electronique et des Technologies du NuméRique (IETR) laboratory, where she leads the VAADER research team. Her research activities deal with computer vision, 3D reconstruction, image and video compression, and representations for 3D videos and multiview videos.
\end{IEEEbiography}
\vfill


%\newpage


%\vfill


\end{document}